\documentclass[10pt, conference, compsocconf, letterpaper]{IEEEtran}
\IEEEoverridecommandlockouts
\pdfoutput=1

\usepackage{graphicx}
\usepackage{graphicx}
\usepackage{mathtools}
\usepackage{braket}
\usepackage{listings}
\usepackage{titlesec}
\usepackage{booktabs}
\usepackage{array}
\usepackage{hyperref}
\usepackage{subcaption}
\newcolumntype{L}[1]{>{\raggedright\let\newline\\\arraybackslash\hspace{0pt}}m{#1}}

\usepackage{latexsym}
\usepackage{qcircuit}

\usepackage{amssymb}
\newcommand{\ctrlb}[1]{\push{\rule{-0.1em}{0em}{\square}\rule{-0.1em}{0em}}  \qwx[#1] \qw} 

\makeatletter
\def\lastoftwo@hyphen#1-#2\@nil{#2}%
\def\submynthofm#1#2\@nil{\lastoftwo@hyphen#2\@nil}%
\newcommand\subfirstoffive[5]{\submynthofm#1\@nil}%
\newcommand\sref[1]{\expandafter\real@setref\csname r@#1\endcsname\subfirstoffive{#1}}
\makeatother
\usepackage{tabularx}
\usepackage{multirow}

\lstdefinelanguage{qasm}
{
  morekeywords={
    qubits,
    name,
    map,
    mov,
    cnot, h, x, y, x, not, cx, cz, cr, crk, prep_x, prep_z, prep_y, rx, ry, rz, s, t, sdag, tdag, measure, display, measure_x, measure_z, measure_parity,version
  },
  sensitive=false, 
  morecomment=[l]{\#}, 
  morecomment=[s]{/*}{*/}
}

\usepackage[table,xcdraw]{xcolor}

\colorlet{punct}{red!60!black}
\definecolor{background}{HTML}{EEEEEE}
\definecolor{delim}{RGB}{20,105,176}
\colorlet{numb}{magenta!60!black}

\lstdefinelanguage{json}{
    basicstyle=\normalfont\ttfamily,
    breaklines=true,
    basicstyle=\scriptsize\ttfamily, 
    captionpos=b, 
    extendedchars=true, 
    tabsize=2, 
    columns=fixed, 
    keepspaces=true, 
    showstringspaces=false, 
    breaklines=true, 
    frame=bt, 
    numbers=left, 
    numberstyle=\tiny\ttfamily, 
    numbersep=-5pt,
    literate=
     *{0}{{{\color{numb}0}}}{1}
      {1}{{{\color{numb}1}}}{1}
      {2}{{{\color{numb}2}}}{1}
      {3}{{{\color{numb}3}}}{1}
      {4}{{{\color{numb}4}}}{1}
      {5}{{{\color{numb}5}}}{1}
      {6}{{{\color{numb}6}}}{1}
      {7}{{{\color{numb}7}}}{1}
      {8}{{{\color{numb}8}}}{1}
      {9}{{{\color{numb}9}}}{1}
      {:}{{{\color{punct}{:}}}}{1}
      {,}{{{\color{punct}{,}}}}{1}
      {\{}{{{\color{delim}{\{}}}}{1}
      {\}}{{{\color{delim}{\}}}}}{1}
      {[}{{{\color{delim}{[}}}}{1}
      {]}{{{\color{delim}{]}}}}{1}
}

\definecolor{eclipseBlue}{RGB}{42,0.0,255}
\definecolor{eclipseGreen}{RGB}{63,127,95}
\definecolor{eclipsePurple}{RGB}{127,0,85}

\def\note#1{}
\def\note#1{\textbf{\color{red}[#1]}}

\lstdefinelanguage{none}{}

\newcommand\Mark[1]{\textsuperscript{#1}}

\usepackage[capitalise]{cleveref}
\usepackage{xfrac} 
\usepackage{float} 
\usepackage{enumitem} 
\usepackage{placeins} 

\usepackage{svg} 

\mathchardef\mhyphen="2D  
\def\noteAK#1{}

\newbox\tempboxa
\setbox\tempboxa=\hbox{} 
\immediate\pdfxform\tempboxa 
\edef\emptyicon{\the\pdflastxform}

\newcommand\annotate[2]{%
\leavevmode
\pdfstartlink user{%
    /Subtype /Text
    /Contents  (#2)
    /AP <<
      /N \emptyicon\space 0 R
    >>
  }%
  #1
  \pdfendlink%
}
\begin{document}

\title{Efficient decomposition of unitary matrices in quantum circuit compilers\\[1.25ex]
}

\author{\normalfont\large A. M. Krol\Mark{\S}, A. Sarkar\Mark{\S}, I. Ashraf\Mark{\P}, Z. Al-Ars\Mark{\S}, K. Bertels\Mark{\ddag} \\
    \newline \\ \vspace{0.2cm}
    \Mark{\S} Quantum \& Computer Engineering Dept., Delft University of Technology Delft, The Netherlands \\
    \vspace{0.2cm}
    \Mark{\P} HITEC University, Taxila, Pakistan \\
    \vspace{0.2cm}
    \Mark{\ddag}  University of Porto, Portugal \\ }

\maketitle

\thispagestyle{plain} 
\pagestyle{plain} 

\begin{abstract}
Unitary decomposition is a widely used method to map quantum algorithms to an arbitrary set of quantum gates. Efficient implementation of this decomposition allows for translation of bigger unitary gates into elementary quantum operations, which is key to executing these algorithms on existing quantum computers. 
The decomposition can be used as an aggressive optimization method for the whole circuit, as well as to test part of an algorithm on a quantum accelerator. For selection and implementation of the decomposition algorithm, perfect qubits are assumed. We base our decomposition technique on Quantum Shannon Decomposition which generates $O(\frac{3}{4}4^n)$ controlled-not gates for an n-qubit input gate. The resulting circuits are up to 10 times shorter than other methods in the field. When comparing our implementation to Qubiter, we show that our implementation generates circuits with half the number of CNOT gates and a third of the total circuit length. In addition to that, it is also up to 10 times as fast. Further optimizations are proposed to take advantage of potential underlying structure in the input or intermediate matrices, as well as to minimize the execution time of the decomposition.
\end{abstract}




\section{Introduction}
Quantum computing is promising to provide the next phase of performance improvement for large scale computing.  
To this end, many different algorithms have been developed in the theoretical domain, such as Shor's algorithm for prime factorization in polynomial time \cite{Shor:1997}, or Grover's algorithm for finding a specific input corresponding to some output in $\sqrt{N}$ time \cite{grover:1997}. 

Recent years have seen some big strides in the field of physical implementations of quantum computers as well. However, these still have some big limitations on the number of qubits, the error rates and the length of the circuits that can be executed on them. Although quantum computers with as many as 128 qubits already exist \cite{art:realizingquantumalgorithms}, error-rates are of the order $10^{-2} - 10^{-3}$ per gate \cite{art:notallqubitsequal}. Therefore, to execute a circuit on a physical quantum chip, requires significant error-correction, as well as mapping, scheduling and other such measures \cite{openql:2020}.


In the meantime, many high-level quantum programming languages are being developed. These can be used to write algorithms for future quantum computers, without the strict limits imposed by the current state of physical qubits. 


These algorithms are executed on simulators, which comes with its own set of restrictions. Some simulators require the use of specific qubit topology, limit possible qubit states or the number of qubits, and all of them are bound by the classical resources of the system the simulation is run on. The main resource limit is the memory necessary to store the quantum circuit and the total qubit state, which is dependent on the length of the circuit, the number of qubits and the degree of superposition. These also influence the processing time necessary to simulate the full circuit, which is generally done by some form of matrix multiplications of the qubit state and each gate in the circuit. 

Unitary Decomposition is the process of translating an arbitrary \textit{unitary}\footnote{A unitary matrix $U$ is a square, complex matrix, of which the inverse ($U^{-1}$) and the conjugate transpose ($U^\dagger$) are the same, i.e. $U^\dagger = U^{-1}$ and $UU^\dagger = I$. \cite{chapter4:unitarymatrices}} gate into a specific (universal) set of single and two-qubit gates. Unitary decomposition is necessary because it is not otherwise possible to execute an arbitrary quantum gate on either a simulator or quantum accelerator. This makes it a required feature for algorithms that use any type of gate that is not supported by the target platform, or just produce an arbitrary unitary gate that will need to be translated. 

This paper proposes a highly-efficient method to implement unitary decomposition for quantum algorithms using the Quantum Shannon Decomposition. The paper shows that our approach is up to 10x more efficient in terms of the number of gates generated for a given unitary matrix size, and requires up to a 100 times less wall-clock execution time than other implementations. The contributions of this paper are as follows:
\begin{itemize}
    \item Implementation of Quantum Shannon Decomposition for unitary decomposition of quantum algorithms.
    \item Decomposition optimizations that take advantage of the underlying matrix structure.
    \item Integration and evaluation of our method in the OpenQL quantum programming framework.
    \item Optimizing the implementation of quantum genome analysis use-case using our method
\end{itemize}

This paper is structured as follows. In \cref{sec:whyunittdecomp}, applications for unitary decomposition are discussed. Then, in \cref{sec:notation}, some background is given on qubits, gate-based computation and the special qubit gates that will be used. The specific decomposition method for multi-controlled gates is given in \cref{sec:multicontrolled}. In \cref{sec:comparison} several decomposition algorithms are compared based on their resulting CNOT-count. 
The implementation of the selected algorithm, Quantum Shannon Decomposition, is outlined in \cref{sec:implementation}. Optimizations to this implementation can be found in \cref{sec:optimization}. Experimental results are shown in \cref{sec:results}, and compared to other implementations in \cref{sec:comparequbiter}. Finally the conclusion and future work can be found in \cref{sec:conclusion}.

\section{Motivation for unitary decomposition} \label{sec:whyunittdecomp}
Unitary decomposition is useful in several contexts. The first is the broad class of algorithms that generate arbitrary unitary gates that need to be translated into a quantum circuit. But also to enable more modular design of quantum algorithms or as an aggressive optimization method. 

We will use two quantum algorithms that we have developed in the context of genome sequencing as an example of a possible application for unitary decomposition. 
With genome sequencing, a genome sequence is first read as many short pieces, which then need to be combined to get the full DNA sequence. This is currently done using many different algorithms, which are executed using (classical) high performance computing systems \cite{art:genomesequencingzaid}.

For genome sequencing using quantum accelerators, the DNA sequences can be stored in superposition. The two algorithms that will be discussed both use a unitary matrix in the process of finding the position of a short read (sequence of a small piece of DNA) on a reference genome. That matrix needs to be decomposed before the algorithm can be run on a quantum accelerator or simulator \cite{art:qibam}.

The first quantum genome sequencing algorithm we will use is Quantum indexed Bidirectional Associative Memory (QiBAM) \cite{art:qibam} (\cref{eq:QiBAMU}), which uses a unitary oracle assembled from a binomial distribution (\cref{eq:QiBAMb}). Here, $\gamma$ is a factor which influences the width of the distribution, $h(p,x)$ is the Hamming distance between the query pattern ($p$) and all memory states ($x$), and $d$ is the number of qubits required to store the memory states. $d$ is also the size of the vector and resulting matrix. 
\begin{align}
    U(2^d) &= I(2^n) - 2 \ket{b_p}\bra{b_p} \label{eq:QiBAMU} \\
    \ket{b_p^x} &= \sqrt{\gamma^{h(p,x)}(1-\gamma)^{d-h(p,x)}} \label{eq:QiBAMb}
\end{align}
The second genome sequencing algorithm is Quantum Associative Memory (QAM). This uses a Hadamard-like transformation to store the patterns, assembled using \cref{eq:QAM} \cite{art:quantumassociativememory}. 

\begin{align}\setlength\arraycolsep{2pt}\renewcommand*\arraystretch{1.2}
\hat{S}^{S_pp} = CR_y(2sin^{-1}(-\sfrac{1}{\sqrt{p}}) = \begin{bmatrix} 1 & 0& 0& 0 \\ 0 & 1 & 0 & 0\\ 0 & 0 & \sqrt{\frac{p-1}{p}} & \frac{-s}{\sqrt{p}} \\ 0 & 0 &  \frac{s}{\sqrt{p}} & \sqrt{\frac{p-1}{p}} \end{bmatrix} \label{eq:QAM}
\end{align}

In order to apply either gate from these two algorithms to qubits, they need to be translated into some combination of (elementary) quantum gates that can be executed on a quantum accelerator. And the same goes for other such algorithms. 

Besides that, unitary decomposition also facilitates short-cuts in the design of new algorithms. With unitary decomposition, a developer can keep part of an algorithm as a unitary gate/matrix while working on some other part and test this. Otherwise, the algorithm can only be executed in full when all of it is made out of known quantum gates.  Unitary decomposition allows the full algorithm to be tested and checked much earlier in the development process on the target quantum chip or simulator.

Furthermore, unitary decomposition can be used as an aggressive optimization method, because the maximum number of gates resulting from a decomposition can be calculated easily beforehand. The maximum length of the circuit resulting from the decomposition is only dependent on the number of qubits affected by the gate. For circuits longer than this maximum, so consisting of more gates, assembly of all gates into a unitary matrix and then decomposing that matrix will always result in a shorter circuit.

Someone programming in OpenQL might, for example, specify a circuit with three qubits with 180 gates, this might be because of application semantics, code-readability or because they did not consider the optimal way to program their quantum algorithm. 180 gates is more than the gates that would result from decomposing an arbitrary 3-qubit gate. So if the circuit is combined into a single unitary matrix and then that matrix is decomposed using Shannon Decomposition for example, then the length of the circuit will have gone from 180 gates to only 120 (84 rotation gates and 36 CNOT gates). 

Something to consider, however, is that the circuit resulting from the decomposition of a unitary matrix is longer than the theoretical minimum. And even the theoretical minimum number of gates for a general n-qubit unitary gate becomes quite large very quickly, since it scales with $4^{n-1}$ in the leading term. So in most cases, a hand-optimized and application specific circuit will be shorter than the one resulting from universal unitary decomposition. But these hand-optimized circuits are labour-intensive and require a significant amount of time to develop, while unitary decomposition can be done automatically.

\section{Background} \label{sec:notation}
In this section, background and notation will be given for qubits, quantum gates, unitary matrices, the universal set of gates that will be used, quantum multiplexers and multi-controlled gates. 

\subsection{Qubit notation}
A qubit state is represented in braket notation as: 
\begin{align}
    \ket{\phi} = \alpha\ket{0} + \beta\ket{1}
\end{align}


Besides the $\ket{\_}$ notation, quantum states can also be represented as complex vectors: $\alpha\ket{0} + \beta\ket{1} = \begin{bmatrix} \alpha & \beta \end{bmatrix}^T$. 
This is especially useful for the combined state of multiple qubits, where the first row of the vector corresponds to the binary number "0" in as many bits as there are qubits. The second row corresponds to the number "1", etc. As an example, for a three qubit state the first row corresponds to $\ket{000}$, and the second to $\ket{001}$. This continues to the final row, which is $\ket{111}$. 
The state vector has $2^n$ rows for the state of $n$ qubits. 

\subsection{Quantum gates}
Qubits are manipulated using gates, which are matrices that operate on the qubit state vector. To calculate the effect of gates on the combined qubit state, the state vector is multiplied by the matrix representations of the gates in reverse order. 


In the circuit notation, each line going into or out of a gate represents one qubit. To represent n-qubit gates, so gates that affect an unspecified number of qubits, a line with a backslash through it is used.

\begin{align*}
(n+1) \text{-qubit gate} =\begin{split}
\Qcircuit @C=0.5em @R=.2em { \push{\rule{0em}{1em}}&\push{1~ qubit~}&\push{\rule{0em}{1em}} &\qw &\multigate{1}{U} &\qw &\\
\push{\rule{0em}{1em}}&\push{n~ qubits}&\push{\rule{0em}{1em}} &{\backslash}\qw &\ghost{U} &\qw &}  \end{split}
\end{align*}

\subsection{Unitary matrices}
The matrix representation of a (pure) quantum gate always corresponds to a unitary matrix, which is why the decomposition method in this paper is called unitary decomposition. 

Unitary matrices are written as $U(2^n)$, which means a $2^n{\times}2^n$ matrix corresponding to an n-qubit gate, which has the following properties\cite{chapter4:unitarymatrices}: 
\begin{itemize}
    \item $U^\dagger = U^{-1}$
    \item $U$ is diagonizable
     \item $|det(U)| = e^{i\theta}$ for any $\theta$ \cite{introductiontoliegroups}
    \item For $U = \begin{bmatrix} A & B \\ C & D\end{bmatrix}$,  $\sqrt{A^2 + B^2} = 1$.
\end{itemize}
The Special Unitary group, $SU$ is a subgroup of unitary matrices where:
\begin{itemize}
    \item $|det(U)| = 1$ for $U$ in $SU$ \cite{introductiontoliegroups}
\end{itemize}
Unitary matrices in the Special Unitary group are written as $SU(2^n)$. 

When a measurement is performed, the global phase ($\Phi$) of the qubits does not influence the measurement probabilities. This means that all quantum gate operations can be represented by a matrix in $SU(2^n)$\cite{art:arbitrarytwoqubit}. 
These properties will be used to decompose the matrix, using one of the algorithms described in \cref{sec:comparison}.

\subsection{Universal set of gates} \label{subsec:universalset}
In order to decompose all possible unitary matrices into quantum gates, a universal gate set is selected. This means the decomposition will result in circuits with (only) the following three gates: rotations around the Z and Y axis by an arbitrary angle, the $R_z(\theta)$ and $R_y(\theta)$ gates, and the controlled not, the CNOT gate. The matrices for these are shown in \cref{eqn:rotyz,eqn:rotz,eqn:cnot}.
\begin{align}
\label{eqn:rotyz}
\begin{split}
R_y(\theta) = \begin{bmatrix} cos\left(\sfrac{\theta}{2}\right) & sin\left(\sfrac{\theta}{2}\right) \\ \mhyphen sin\left(\sfrac{\theta}{2}\right) & cos \left(\sfrac{\theta}{2}\right)\end{bmatrix}
\end{split}
\begin{split}
=~~
\end{split}
\begin{split}
\Qcircuit @C=0.5em @R=.7em {
&\gate{R_y}&\qw
}
\end{split} \\[2pt]
\begin{split}
R_z(\theta) =\begin{bmatrix} e^{\mhyphen i\theta /2} & 0\\ 0 & e^{i\theta/2}\end{bmatrix} \label{eqn:rotz}
\end{split}
\begin{split}
= ~~
\end{split}
\begin{split}
\Qcircuit @C=0.5em @R=.7em {
&\gate{R_z}&\qw
}
\end{split} \\[2pt]
\begin{split}
\textit{CNOT} = \begin{bmatrix} 1 & 0 & 0 & 0 \\ 0 & 1 & 0 & 0 \\ 0 & 0 & 0 & 1 \\ 0 & 0 & 1 & 0\end{bmatrix} \label{eqn:cnot}
\end{split}
\begin{split}
=~~
\end{split}
\begin{split}
\Qcircuit @C=0.5em @R=1em {
&\push{\rule{0em}{0em}}&\qw &\ctrl{1}&\qw &\qw \\
&\push{\rule{0em}{0em}}&\qw &\targ &\qw &\qw 
}
\end{split}
\end{align}%
\subsection{Quantum multiplexers} \label{subsec:quantumMultiplexers}
Besides these conventional gates, there are several gates used in this paper as intermediate results for the various decomposition algorithms.

The first is the quantum multiplexer, which corresponds to a unitary matrix corresponding with the following structure~\cref{eq:ucagmat}. 
\begin{align}
    &U(2^n) = \begin{bmatrix} U_0(2^{n-1}) & 0 \\0 & U_1(2^{n-1}) \end{bmatrix}\label{eq:ucagmat}
\end{align}
Here, $U(2^n)$ denotes a unitary gate over $n$ qubits, which is a unitary matrix of $2^n$ rows and $2^n$ columns.  $U_0(2^{n-1})$ and $U_1(2^{n-1})$ are both (n-1)-qubit gates. The rest of the matrix of $U$ is zero. The gate is uniformly controlled, which means that when the control is 0, the upper left ($U_0$) of the matrix affects the qubits. But when the control is 1, the lower right gate ($U_1$) gets applied. 
In a circuit, this looks like:
\begin{figure}[H]
    \centering
    \leavevmode
\annotate{
\Qcircuit @C=0.5em @R=.7em {
&\lstick{1}&\qw &\multigate{1}{U(2^n)}   &\qw    &
& &\qw &\ctrlb{1}    &\qw        & \\
&\lstick{n-1}&\backslash\qw  &\ghost{U(2^n)}          &\qw    &\push{\rule{.3em}{0em}\raisebox{2.2em}{=}\rule{.3em}{0em}} 
& &\backslash\qw & \gate{U_0~or~U_1}  &\qw  }
}{A quantum circuit equivalence with on the left of the equal sign a multi-qubit gate with two inputs, 1 qubit and (n-1)-qubits. The (n-1)-qubits are shown as a line with a backslash through it. On the right of the equal sign is the quantum multiplexor: it is shown as a regular gate applied to the bottom qubits with "U0 or U1" in it. This gate is connected to a box-control on the top qubit, which is the control of the quantum multiplexer.}
\end{figure}
The first line is the controlling qubit, and the lower line is the rest (n-1) of the qubits. The box with the line to the lower gate means that it is uniformly controlled. 

\subsection{Multi-controlled (rotation) gates} \label{subsec:multicontrolledgateS}
Another common intermediate gate is the multi-controlled (rotation) gate. This is a 1-qubit gate with $k$ control bits. Rather than just applying a gate when all control bits are zero, the applied operation to the target qubit can be different for each of the $2^k$ possible classical values of the control qubits.

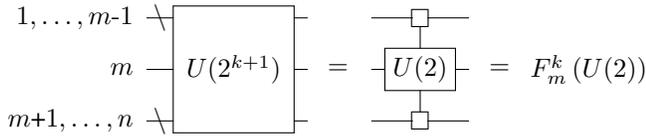
\begin{figure}[H]
\leavevmode
\hfill \annotate{\Qcircuit @C=0.5em @R=.7em {
&\lstick{ 1,\dotsc ,m\mhyphen1}&\backslash\qw &\multigate{2}{U(2^{k+1})} &\qw & &
& \ctrlb{1}   &\qw    
&&\\
&\lstick{m}          &\qw   &\ghost{U(2^{k+1}) }     &\qw & \push{\rule{0.0em}{0em}=\rule{0.1em}{0em}}&
& \gate{U(2)}   &\qw 
& \push{\rule{0.0em}{0em}=\rule{0.0em}{0em}}& \push{F_m^{k}\left(U(2)\right)}\\
&\lstick{m\textrm{+}1,\dotsc,n} &\backslash\qw&\ghost{U(2^{k+1})}       &\qw & &
& \ctrlb{-1}   &\qw    
&&}
}{A quantum circuit equivalence with on the left of the equal sign a multi-qubit gate with three inputs. The first is labeled 1,... ,m-1, it represents the first m-1 qubits. To show this, the input has a backslash through it. The middle input is labeled m, for qubit number m. The last input is labeled m+1 ,... ,n, it represents the qubits from m+1 to n. Inside the multi-qubit gate is written U(2 to the power of (k+1)). On the right of the equal sign is a quantum circuit with three lines, which are the same as the inputs of the multi-qubit gate. A U(2) gate is applied to the middle qubit, and it is connected to square control boxes on the top and bottom lines. This shown to be equal to F\^k\_m(U(2)).}
\caption{A multi-controlled U(2) gate}
\label{fig:fullycontrolled}
\end{figure}

This is written as $F^k_m(U(2))$, which is a fully or multi-controlled U(2) gate with $k$ control qubits, with the target qubit at position $m$. The circuit representation of this gate is shown in \cref{fig:fullycontrolled}. To indicate that an operation is applied for either state of the control bits, a square control box is used. 

These multi-controlled gates correspond to a (block) diagonal unitary matrix, which is why they show up frequently in decomposition schemes. This is shown in \cref{eq:multicontrolledU2}.

\begin{align}
    F^k_m(U(2)) = diag_j \left( U(2)_j \right) = \setlength\arraycolsep{2pt}
    \begin{bmatrix} 
        U(2)_0 & & \\
               & \ddots & \\
               & & U(2)_{2^k}  \end{bmatrix} \label{eq:multicontrolledU2}
\end{align}
A multi-controlled rotation gate around axis $a$ corresponds to the matrix shown in \cref{eq:multicontrolledRa}. This can be any axis, but in the paper mainly the multi-controlled $R_y$ and $R_z$ will be used.

\begin{align}
    F^k_m(R_a) = diag_j \left( R_a(\theta_j) \right) = \setlength\arraycolsep{2pt}
    \begin{bmatrix} 
        R_a(\theta_0) & & \\
               & \ddots & \\
               & &R_a(\theta_{2^k}) \end{bmatrix} \label{eq:multicontrolledRa}
\end{align}




\section{Decomposing multi-controlled rotation gates} \label{sec:multicontrolled}
\begin{figure}[H]
\centering
\leavevmode
\annotate{
\Qcircuit @C=0.4em @R=.7em {
&\qw            & \ctrlb{1}     &\qw        &           &\push{\rule{1.5em}{0em}}     &   
&\qw            & \qw           &\ctrl{2}   & \qw       & \ctrl{2}          &\qw\\
&\backslash \qw & \ctrlb{1}     &\qw        &\rstick{=} &                   &
&\backslash \qw & \ctrlb{1}     &\qw        & \ctrlb{1} &\qw                &\qw\\
&\qw            & \gate{R_a}    &\qw        &           &           &                   
&\qw            & \gate{R_a}    &\targ      &\gate{R_a} &\targ              &\qw
}}{A quantum circuit that shows a multi-controlled rotation gate with k control bits is equal to a circuit consisting of 2 multi-controlled rotation gates with (k-1) controls and two CNOTs, with first a multi-controlled rotation gate applied to the lower k qubits, then a CNOT applied to the lowest qubit, controlled from the first qubit. Then the second multi-controlled rotation gate and the second CNOT, in exactly the same way.}
\caption{Partial decomposition of an $F^m_k(R_a)$-gate.} \label{fig:multicontrolledRa}
\end{figure}
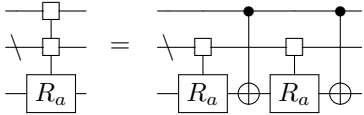

The multi-controlled rotation gates from \cref{subsec:multicontrolledgateS} can be decomposed into a combination of CNOTs and regular rotation gates. This can be done using the method from \cite{art:quantumcircuitsforgeneralmultiqubit}, which results in $2^k$ CNOTs gates and $2^k$ 1-qubit rotation gates for a controlled rotation gate with $k$ control bits. To get from an $F^m_k(R_a)$-gate to an $F^m_{k-1}(R_a)$-gate, a circuit like \cref{fig:multicontrolledRa} can be used.

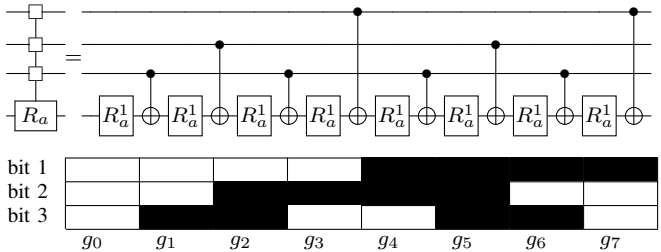
\begin{figure}[H]
\centering
\leavevmode
\footnotesize
\annotate{
\Qcircuit @C=0.4em @R=.7em {
& \ctrlb{1}   &\qw    &\push{\rule{0em}{1em}}
& &\qw 
& \qw & \qw & \qw & \qw & \qw & \qw & \qw & \ctrl{3}
& \qw & \qw & \qw & \qw & \qw & \qw & \qw & \ctrl{3}&\qw\\
&\ctrlb{1}    &\qw    &
& &\qw 
& \qw & \qw & \qw & \ctrl{2} & \qw & \qw & \qw & \qw
& \qw & \qw & \qw & \ctrl{2} & \qw & \qw & \qw & \qw &\qw\\
& \ctrlb{1}   &\qw    &\ustick{=}
& &\qw 
& \qw & \ctrl{1} & \qw & \qw & \qw & \ctrl{1} & \qw & \qw
& \qw & \ctrl{1} & \qw & \qw & \qw & \ctrl{1} & \qw & \qw&\qw\\
& \gate{R_a}   &\qw    &
& &\qw 
& \push{\framebox[1.2\width]{$R_a^1$} } \qw&\targ &\push{\framebox[1.2\width]{$R_a^1$} }\qw&\targ& \push{\framebox[1.2\width]{$R_a^1$} }\qw&\targ& \push{\framebox[1.2\width]{$R_a^1$} }\qw&\targ
& \push{\framebox[1.2\width]{$R_a^1$} } \qw&\targ& \push{\framebox[1.2\width]{$R_a^1$} } \qw&\targ&\push{\framebox[1.2\width]{$R_a^1$} }\qw &\targ&\push{\framebox[1.2\width]{$R_a^1$} } \qw&\targ&\qw\\
} 
\vspace{1em}
\newcolumntype{R}{>{\raggedright\arraybackslash}X}%
\begin{tabularx}{\linewidth}{RXXXXXXXX}
\cline{2-9}
\multicolumn{1}{l|}{bit 1} & \multicolumn{1}{l|}{} & \multicolumn{1}{l|}{}                         & \multicolumn{1}{l|}{}                                                & \multicolumn{1}{l|}{}                                                & \multicolumn{1}{l|}{\cellcolor[HTML]{000000}{\color[HTML]{343434} }} & \multicolumn{1}{l|}{\cellcolor[HTML]{000000}}                        & \multicolumn{1}{l|}{\cellcolor[HTML]{000000}} & \multicolumn{1}{l|}{\cellcolor[HTML]{000000}} \\ \cline{2-9} 
\multicolumn{1}{l|}{bit 2} & \multicolumn{1}{l|}{} & \multicolumn{1}{l|}{}                         & \multicolumn{1}{l|}{\cellcolor[HTML]{000000}{\color[HTML]{343434} }} & \multicolumn{1}{l|}{\cellcolor[HTML]{000000}}                        & \multicolumn{1}{l|}{\cellcolor[HTML]{000000}}                        & \multicolumn{1}{l|}{\cellcolor[HTML]{000000}}                        & \multicolumn{1}{l|}{{\color[HTML]{EFEFEF} }}  & \multicolumn{1}{l|}{}                         \\ \cline{2-9} 
\multicolumn{1}{l|}{bit 3} & \multicolumn{1}{l|}{} & \multicolumn{1}{l|}{\cellcolor[HTML]{000000}} & \multicolumn{1}{l|}{\cellcolor[HTML]{000000}}                        & \multicolumn{1}{l|}{\cellcolor[HTML]{FFFFFF}{\color[HTML]{C0C0C0} }} & \multicolumn{1}{l|}{}                                                & \multicolumn{1}{l|}{\cellcolor[HTML]{000000}{\color[HTML]{343434} }} & \multicolumn{1}{l|}{\cellcolor[HTML]{000000}} & \multicolumn{1}{l|}{{\color[HTML]{EFEFEF} }}  \\ \cline{2-9} 
                           & $g_0$                 & $g_1$                                         & $g_2$                                                                & $g_3$                                                                & $g_4$                                                                & $g_5$                                                                & $g_6$                                         & $g_7$                                                                 
\end{tabularx}
}{The decomposition of a multi-controlled R\_a gate is shown. Each qubit is shown as a separate line, and the decomposition is as follows. The circuit consists of 8 sets of alternating R\^1\_a gates applied to the lowest qubit, and CNOT gates of which the target is also the lowest qubit and the control one of the three upper qubits. Below the figure a table that shows Gray code is shown, with each column as wide as one rotation gate and one CNOT. The 3-bit Gray code is shown as black and white cells in the 3 row table, and which qubit is the control qubit for the CNOT in the quantum circuit is the same number as which bit changes in the Gray code below that CNOT.}
\caption{Decomposition of an $F^3_4(R_a)$-gate}
\label{fig:multicontrolled3}
\end{figure}

This can be extended until only CNOT gates and 1-qubit rotation gates are left, which leads to an example decomposition of a rotation gate with 3 control bits as shown in \cref{fig:multicontrolled3}.

To directly calculate which qubit is the control bit for each CNOT, can be determined using Gray code. This is shown in the table below the circuit. The number of the bit that gets changed in the Gray code is the number of the qubit that will be the control bit. 


For each control bit of the multi-controlled gate, a 1-qubit rotation gate and a single CNOT is used, so for the total decomposition of an $F^k_m$-gate requires $2^k$ rotation gates and CNOTs \cite{art:quantumcircuitsforgeneralmultiqubit}. This is the least-known number of gates for decomposing such a matrix, and is therefore used in almost all decomposition methods for (block) diagonal matrices of this form.
\section{Comparison of different decomposition methods} \label{sec:comparison}
In this section, first the selection criteria for the various decomposition methods will be outlined in \cref{subsec:selectioncriteria}. Then, the theoretical lower bounds for the number of gates resulting from decomposition is given in \cref{subsec:theoreticallowerbound}, with implementations for a 1- and 2-qubit gate in \cref{subsec:zyz,subsec:decompoftwoqubitgates}. This is followed by an examination of various general decomposition methods from literature in \cref{subsec:theorem9,subsec:effdecomp,subsec:csdandnq,subsec:qsd} and finally the selection in \cref{subsec:selection}.

\subsection{Selection criteria} \label{subsec:selectioncriteria}
Quantum computers are currently limited by the error-rates and decoherence of qubits\cite{art:notallqubitsequal}. And the longer the circuit, the higher the chance of errors will become. So therefore the selection will be based on circuit length, although the decomposition algorithm will for now only be tested with perfect qubits on a simulator.


For all decomposition methods, the number of gates resulting from the decomposition is only dependent on the number of qubits affected by the unitary gate. So for generic n-qubit unitary gates, the resulting circuit length can be calculated from the size of the input matrix.

To measure the length of the resulting circuit, the number of CNOT gates will be used. There are several reasons for that.
The first is that not all papers distinguish between generic 1-qubit gates and rotation gates. The decomposition of a generic 1-qubit gate takes three rotation gates (see Section~\ref{subsec:zyz}) so the comparison might be a factor 3 off if 1-qubit gates are used to judge circuit length. The CNOT gate is used as the result for all decomposition methods, and always has the same definition. This makes it a good metric for the total circuit length.

Secondly, each CNOT can generate entangled states between qubits \cite{art:entanglement20qubit}. And for execution of the circuit on (near-term) quantum devices, each CNOT between non-neighboring qubits might introduce additional mapping operations\cite{openql:2020}. So to reduce mapping in the future, a circuit with as few CNOTs as possible is desired.

Thirdly, the error-rates for two-qubit gates are currently considerably higher than for 1-qubit gates \cite{art:notallqubitsequal}. So the chance that an error occurs in a circuit becomes much bigger with more CNOTs. So to make the decomposition feasible for near-term quantum applications, it is not only important to keep the circuit-length low, but especially the CNOT count. 

\subsection{Theoretical lower bounds} \label{subsec:theoreticallowerbound}
There is a theoretical lower bound for the number of CNOTs resulting from the decomposition of an n-qubit gate, and it is mathematically proven to be $\frac{1}{4}(4^n-3n-1)$ \cite{art:minimaluniversal}. There are implementations that reach this number for 1 and 2 qubit gates \cite{art:minimaluniversal}, which will be outlined in the next sections. This lower limit is included in the comparison, because it is useful to keep in mind what is and is not possible in terms of algorithms for unitary decomposition.

\subsection{ZYZ decomposition} \label{subsec:zyz}
For a 1-qubit gate, no CNOT gates are necessary. And if rotation gates around any axis are possible, only one such gate is needed to apply any 1-qubit operation. But when using standard elementary gates, such as rotations around the Pauli X, Y or Z-axis, the decomposition of an arbitrary 1-qubit gate results in 3 rotation gates using ZYZ decomposition~\cite{art:minimaluniversal}. 

One way to do this is with two rotation-z gates and one rotation-y gate. For this decomposition, the angles $\Phi, \alpha, \beta, \gamma$ can be found so that the following equation is satisfied:

\begin{align}
U(2) &= e^{-i\Phi}  \begin{bmatrix} A & B \\ C & D\end{bmatrix} &&= e^{-i\Phi}R_z(\alpha)R_y(\beta)R_z(\gamma)\\
SU(2) &= \begin{bmatrix} A & B \\ C & D\end{bmatrix} &&= R_z(\alpha)R_y(\beta)R_z(\gamma)
\end{align}

\begin{figure}[H]
    \centering
    \leavevmode
    \annotate{
\Qcircuit @C=1em @R=.7em {
&\gate{U} & \qw &\push{\rule{.3em}{0em}=\rule{.3em}{0em}}&  & \gate{R_z(\theta_0)}& \gate{R_y(\theta_1)} & \gate{R_z(\theta_2)} &\qw
}
}{A quantum circuit equivalence is shown. On the left side of the equal sign is a single qubit gate, which is called U. On the right of the equal sign are three 1-qubit gates applied to the same qubit in succession. The first is an R\_z(theta 0) gate, the second an R\_y(theta 1) gate and the third an R\_z(theta 2) gate.}
    \caption{Minimal universal quantum circuit for a 1-qubit gate \cite{art:minimaluniversal}.} 
    \label{fig:minimaluniversalzyz}
\end{figure}
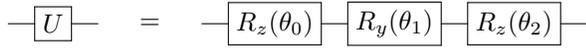

These angles can be calculated using the eigenvalues of the matrix, and are used in the circuit shown in \cref{fig:minimaluniversalzyz}. This is a universal decomposition for a 1-qubit $SU(2)$ gate \cite{art:minimaluniversal}.

\subsection{Minimal decomposition of 2-qubit gates} \label{subsec:decompoftwoqubitgates}
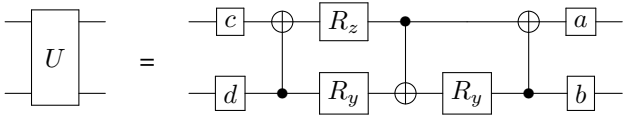
\begin{figure}[H]
    \centering
    \leavevmode
    \annotate{
\Qcircuit @C=1em @R=.7em {
&\multigate{1}{U} & \qw & & &\gate{c} & \targ & \gate{R_z}&  \ctrl{1} & \qw & \targ &\gate{a} &\qw\\
&\ghost{U}&\qw &\push{\rule{.3em}{0em} \raisebox{2.2em}{=}\rule{.3em}{0em}}   & &\gate{d} & \ctrl{-1} &\gate{R_y} &  \targ & \gate{R_y} & \ctrl{-1}&\gate{b}&\qw
}
}{A circuit equivalence is shown, with on the left side of the equal sign a 2-qubit gate, which is called U. On the right of the equal sign is a quantum circuit with 2 qubits. This circuit consists of the following gates, from left to right: a 1-qubit gate called c on the top qubit and a 1-qubit gate called d on the bottom qubit. A CNOT where the target is the top qubit, and the bottom qubit the control. An R\_z gate on the top qubit and an R\_y gate on the bottom qubit. A CNOT where the target is the bottom qubit and the top qubit is the control. No gate on the top qubit and an R_y gate on the bottom qubit. A CNOT where the target is the top qubit. And a 1-qubit gate called a on the top qubit and a 1-qubit gate called b on the bottom qubit. This brings the total to 3 CNOTs, 3 rotation gates and 4 1-qubit gates called a, b, c and d.}
    \caption{Minimal universal quantum circuit for a 2-qubit gate using 18 elementary gates \cite{art:minimaluniversal}.} 
    \label{fig:minimaluniversal}
\end{figure}

From the theoretical lower bounds we know that at least 2.25 CNOT gates are needed for a 2-qubit gate. This rounds up to 3 CNOTs, and a circuit that achieves that number is shown in \cref{fig:minimaluniversal} ~\cite{art:minimaluniversal}.

To obtain the values for the gates of this circuit, first angles $\alpha$, $\beta$ and $\delta$ are found as in the ZYZ decomposition (\ref{subsec:zyz}). These are used to make circuit $v$ so that:
\begin{align}
    (a \oplus b) v (c \oplus d) = U(4) 
\end{align}
With $v$ as the circuit:
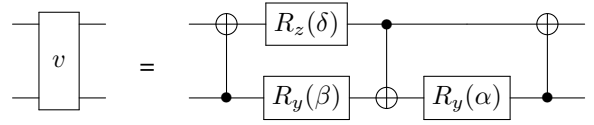
\begin{figure}[H]
    \centering
    \annotate{
\Qcircuit @C=1em @R=.7em {
&\multigate{1}{v}   &\qw & 
&  & \targ      & \gate{R_z(\delta)}    &  \ctrl{1} & \qw               & \targ     &\qw\\
&\ghost{v}          &\qw &\push{\rule{.3em}{0em} \raisebox{2.2em}{=}\rule{.3em}{0em}}   
&  & \ctrl{-1}  &\gate{R_y(\beta)}      &  \targ    &\gate{R_y(\alpha)} & \ctrl{-1} &\qw
} 
}{A circuit equivalence is shown, with on the left side of the equal sign a 2-qubit gate, which is called v. On the right of the equal sign is a quantum circuit with 2 qubits. This circuit consists of the following gates, from left to right: a CNOT where the target is the top qubit, and the bottom qubit the control. An R\_z(delta) gate on the top qubit and an R\_y(beta) gate on the bottom qubit. A CNOT where the target is the bottom qubit and the top qubit is the control. No gate on the top qubit and an R\_y(alpha) gate on the bottom qubit. And a CNOT where the target is the top qubit. This brings the total to 3 CNOTs and 3 rotation gates.}
\caption{The circuit $v$, used to construct a universal 2-qubit gate\cite{art:minimaluniversal}.} 
    \label{fig:minimaluniversal_vcircuit}
    \end{figure}
Then, to get the 1-qubit gates, first matrix $A \in SO(4)$ can be found so that $A U U^T A^\dagger$ is diagonal\footnote{$SO(n)$ is the Special Orthogonal group, which means that the inverse of a matrix $Q$ is  equal its transpose: $Q^{-1} = Q^T$ and $det(Q) = 1$}. Through more diagonalization, $B \in SO(4)$ can be found so $A U U^T A^T = B v v^T B^T$ and matrix $C$ as $C = v^\dagger B^T A U \in SO(4)$.  
This leads to $A^T B v C = U$, and because $A$, $B$ and $C$ are in the special orthogonal group they can be implemented by two unitary gates. After combining $A^T$ and $B$, the four gates can be found as: \cite{art:minimaluniversal}
\begin{align}
    A^T B = a \oplus b\\
    C = c \oplus d
\end{align}
Which gives the circuit in \cref{fig:minimaluniversal}. The four 1-qubit gates can be implemented by three rotation gates each, through ZYZ decomposition, so that the total rotation count is $4\cdot 3 + 3$ and the total CNOT count is just the ones for the circuit $v$, so 3. This matches the theoretical lower bounds for an arbitrary 2-qubit gate. 

\subsection{Decomposition through unentangeling of qubits} \label{subsec:theorem9}
The first of the general decomposition methods uses consecutive unentangling of qubits, since one of the ways to specify an n-qubit gate is by its effect on the base vectors. This technique from \cite{synthesisofquantumlogiccircuits} implements the correct behavior for each vector iteratively using a method similar to QR decomposition, which leaves previous assessed vectors preserved. 

\begin{figure}[H]
    \centering
    \leavevmode
\annotate{
\Qcircuit @C=0.5em @R=.7em {
&\dstick{\ket{\psi}} & &\backslash\qw & \ctrlb{1} & \ctrlb{1} & \qw & \rstick{\ket{\psi'}}\\
& \push{\rule{2em}{0em}} & & \qw & \gate{R_z} & \gate{R_y} & \qw & \rstick{\ket{0}}
}
}{A quantum circuit with two circuit lines. The top is an n-qubit line with a backslash through it, the bottom line represents a single qubit. At the start of this circuit, the total qubit state is labelled phi. The circuit consists of two multi-controlled rotation gates. Both are applied to the bottom qubitm, and are connected to a square control box on the top qubit line. The first is a multi-controlled R\_z gate and the second a multi-controlled R\_y gate. After these gates, the top qubit line is labelled phi' and the bottom as qubit state 0.}
\caption{Unentangling a qubit state \cite{synthesisofquantumlogiccircuits}.} \label{circ:theorem9}
\end{figure}
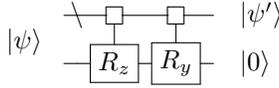

%
To get here, the qubit state vector is divided into $2^{n-1}$ vectors of each 2 elements, which are labeled $\ket{\psi_j}$ for $j = 0, \dotsc, 2^{n-1}-1$. For each of these vectors, \cref{eq:theorem9eq9} is used to determine $r_j$, $t_j$, $\phi_j$ and $\theta_j$. 
\begin{align}  
    &\ket{\psi} = re^{it/2}\left[ e^{\mhyphen i\phi /2} cos \frac{\theta}{2}\ket{0} + e^{i\phi/2}sin\frac{\theta}{2}\ket{1} \right] \label{eq:theorem9eq9}
\end{align}
So that:
\begin{align}
   &R_z(-\phi_j)R_y(-\theta_j)\ket{\psi_j} = r_j e^{it_j}\ket{0} \label{eq:theorem91}
\end{align}
This corresponds to a circuit with a multi-controlled $R_y$ and $R_z$, which is used to unentangle the last qubit. 
The new qubit vector is assembled as in \cref{eq:psinew}. The circuit to translate $\ket{\phi}$ into $\ket{\phi'}\ket{0}$ will be called $E_k$ as in \cref{eq:theorem9total}. 

\begin{align}
  &\ket{\psi'} = \sum_{j=0}^{2^{n-1}-1} r_j e^{it_j}\ket{j} \label{eq:psinew} \\
  &F^{n-1}_n(R_y) F^{n-1}_n(R_z) \ket{\psi} = E_k \ket{\psi} = \ket{\psi'}\ket{0}  \label{eq:theorem9total}
\end{align}
This circuit is implemented with the multi-controlled $R_y$ and $R_z$ gate, as is shown in \cref{circ:theorem9}. 

This method can be applied recursively to map an n-qubit state $\phi$ to a scalar multiple of a bit-string $\ket{b}$, which uses  $2^{n+1}-2n$ CNOT gates. 

Using this method to decompose unitary gates requires $2^{n-1}$ of these state preparation steps. At each step, a circuit $C_j$ is found that maps a unitary gate $U_j$ to a scalar multiple of $\ket{j}$ so that $U_{j+1}= C_j U_j$. 
The final product $U_{2^n-1}$ will be a diagonal gate D, which can be implemented with a multi-controlled $R_Z$ gate, so that $U(2^n) = C_0^\dagger C_1^\dagger \cdots C_{2^n-2}^\dagger D$. 

Each circuit $C_j$ needs ($2^{n+1}-n$) CNOT gates, and with the final diagonal gate this leads to a total of $2\cdot 4^n - (2n+3)\cdot2^n+2n$ CNOT gates \cite{synthesisofquantumlogiccircuits}.

\subsection{Decomposition with Givens rotations} \label{subsec:effdecomp}
In \cite{art:efficientdecompositionofquantumgates} a method of decomposition is described that uses the Givens rotation matrices to do QR factorization of a unitary matrix. 
Each Givens rotation nullifies the element on the $i^th$ column and $j^th$ row of a $U(2^n)$ matrix, as:
\begin{align}
    ^1 G_{n,n-1} U &=  \begin{bmatrix} 
                u_{1,1}     & u_{1,2}   & {\displaystyle \cdots }   & u_{1,n}      \\[-.4em]
                \vdots      & \vdots    & \ddots                    & \vdots \\[0.1em]
                u_{n-2,1}   & u_{n-2,2} & {\displaystyle \cdots}    & u_{n-2,n}   \\[.4em]
                \tilde{u}_{n-1,1}   & \tilde{u}_{n-1,2} & {\displaystyle \cdots}    & \tilde{u}_{n-1,n}   \\[.4em]
                0           & \tilde{u}_{n,2} & {\displaystyle \cdots}    & \tilde{u}_{n,n}   
                \end{bmatrix} \label{eq:givensU}
\end{align}%
The modified elements of U are indicated with a tilde, and the element on the lower left $u_{n,1}$ is nullified by the Givens rotation. 
Each Givens rotation matrix is equal to the identity matrix except for $c=cos(\theta)$ and $s=sin(\theta)$ for the elements at positions $(i,i)$, $(i,j)$, $(j,i)$ and $(j,j)$, with $\theta$ the angle of the Givens rotation. 
These are to nullify elements until all elements below the diagonal are zero, at which point the following equality holds: \cite{art:efficientdecompositionofquantumgates}

\begin{align}
    \left(\prod_{i=1}^{2^{n}-1} \prod^{2^{n}}_{j=i+1} ~^{i} G_{j,j-1} \right) U = I
\end{align}

\begin{figure}[H]
    \centering
    \leavevmode
\annotate{
\Qcircuit @C=0.5em @R=.7em {
&\multigate{1}{U} & \qw & & 
& \ctrlo{1} & \gate{^2\Gamma^\dagger_{4,2}} & \ctrlo{1} & \ctrl{1} & \gate{^1\Gamma^\dagger_{4,2}} & \ctrlo{1} & \qw \\
&\ghost{U}&\qw &\push{\rule{.3em}{0em} \raisebox{2.2em}{=}\rule{.3em}{0em}}   &
& \gate{^3\Gamma^\dagger_{3,4}} & \ctrlo{-1} & \gate{^2\Gamma^\dagger_{3,4}} & \gate{^1\Gamma^\dagger_{2,1}}& \ctrlo{-1} & \gate{^1\Gamma^\dagger_{3,4}} & \qw
}
}{A circuit equivalence is shown, with on the left side of the equal sign a 2-qubit gate, which is called U. On the right of the equal sign is a quantum circuit with 2 qubits. This circuit consists of 6 controlled Gamma gates, applied to the top or bottom qubit, and controlled by the other qubit. All the gates except one are connected to an open circle control. The gates are in order: a Gamma gate on the bottom qubit, a Gamma gate on the top qubit, a Gamma gate on the bottom qubit, another Gamma gate applied to the bottom qubit, but this one with a regular control, and a Gamma gate on the top qubit.}
    \caption{Decomposition into the Givens rotations \cite{art:efficientdecompositionofquantumgates}.}
     \label{circ:givensrotations}
\end{figure}
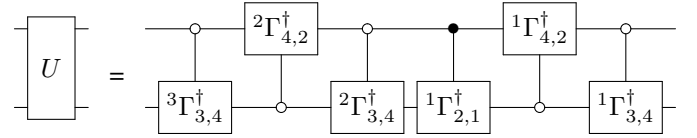

By reordering the base vectors according to Gray code (see \cref{subsec:multicontrolledgateS}), the cosine and sine elements will all be adjacent. This is convenient for quantum computation, because that means that each Givens rotation matrix can be implemented by a controlled 1-qubit gate, $C^k_i$, with $k$ control bits. For one specific combined state of the control qubits the $\Gamma$ gate gets applied to qubit $i$, while for all other states the target qubit is left unaffected. So that:
\begin{align}    
    &\prod_{i=1}^{2^n-1} \prod_{j=i+1}^{2^n} C^{n-1}_{\gamma(i)}(\Gamma^\dagger_{(j,k)}) = SU(2^n) \\
    ^i\Gamma_{j,k} &:= \begin{bmatrix} ^ig_{k,k} & ^ig_{k,j} \\ ^ig_{j,k} & ^ig_{j,j}\end{bmatrix} \label{eq:igamma}
\end{align}%
where $\gamma(i)$ denotes the $i^{th}$ number of the Gray code and the gates $^i\Gamma_{j,k}$ are formed from the matrix for the Givens rotations.
This results in the circuit shown in \cref{circ:givensrotations} for the {decomposition} of a 2-qubit gate.
%

The number of elementary gates and CNOTs were calculated using \cite{art:elementary}, which are the numbers included in the table. 
Generally, this decomposition requires approximately $8.4 \cdot 4^n$ controlled gates, which follows from a recursive relation of $g_n(k) = g^0_n(k) + g_{n-1}(k) + g_{n-1}(k-1)$ \cite{art:efficientdecompositionofquantumgates}.

\subsection{Recursive CSD} \label{subsec:csdandnq}
With the circuit presented in \cite{misc:rudimentaryquantumcompiler}, an n-qubit gate is decomposed into multi-controlled rotation gates. CSD is applied recursively until all the matrices are diagonal.
 
With CSD, any even-dimensional unitary matrix U can be decomposed into real diagonal matrices C and S, and smaller unitary matrices $L_0$, $L_1$, $R_0$, $R_1$ as shown in \cref{eq:CSD0}\cite{art:historyandgeneralityofthecsdecomposition}.
\begin{align}\renewcommand*{\arraystretch}{1.2}
    U = \begin{bmatrix}U_{00}&U_{01}\\U_{10}&U_{11}\end{bmatrix}= \begin{bmatrix}R_0& 0\\ 0& R_1\end{bmatrix} \begin{bmatrix}C&-S\\S&C\end{bmatrix} \begin{bmatrix}L_0&0 \\ 0& L_1\end{bmatrix}\label{eq:CSD0}
\end{align}
The left and right matrices are uniformly controlled gates, see \cref{subsec:quantumMultiplexers}. $C$ and $S$ are diagonal matrices with as the diagonal elements respectively the cosines and sines of angles $\theta_j$ between the subspaces, as shown in \cref{eq:diagC,eq:diagS}. 
\begin{align}
    &C = diag(cos(\theta_0),\dotsc,cos(\theta_n)) \label{eq:diagC}\\
    &S = diag(sin(\theta_0),\dotsc,sin(\theta_n)) \label{eq:diagS}
\end{align}
where the values $\theta$ are ordered from large to small, and are between $\frac{1}{2}\pi$ and 0.

The central matrices from each recursive step correspond to multi-controlled $R_y$ gates which are decomposed as in \cref{subsec:multicontrolledgateS}. The other diagonal gates can be decomposed into a circuit consisting of $\sfrac{1}{2}\cdot n\cdot4^n-\sfrac{1}{2}\cdot2^n$ CNOTs and $\sfrac{3}{2}\cdot4^n - \sfrac{1}{2} \cdot 2^{n}$ 1-qubit rotation gates\cite{misc:qubitergithub}.

This is significantly improved upon in \cite{art:decompositionofgeneralquantumgates}, which stops the recursion at uniformly controlled 1-qubit gates. 
Furthermore, it proves that any uniformly controlled 2-qubit gate ($F_n^{n-1}(U(2))$) can be decomposed into a specific sequence of $2^{n-1}-1$ CNOT gates,  $2^{n-1}$ 1-qubit gates and one total global phase gate expressed as $\Delta_{n}$. 

Furthermore, it proves that each multi-controlled 2-qubit gate can be decomposed into a diagonal gate ($\Delta$) and a Gray code sequence of CNOTs and 1-qubit gates. The diagonal gates are folded into the central matrix from the CSD, so the total decomposition is:
\begin{align}
    & U = \Delta_n \tilde{F}_n^{n\mhyphen1}(U(2))\prod_{i=1}^{2^{n \mhyphen1}\mhyphen1} \tilde{F}_{n-\gamma(i)}^{n\mhyphen1}(U(2))\tilde{F}_{n}^{n\mhyphen1}(U(2)) \label{eq:csdworse}
\end{align}
Each $\tilde{F}_{n}^{n\mhyphen1}(U(2))$ is decomposed with $2^{n-1}-1$ CNOTs, and the $\Delta_n$ gate is implemented with multi-controlled $R_Z$ gates. This results in $2^n-2$ CNOTs, which makes the total CNOT count $\sfrac{1}{2}\cdot4^n - \sfrac{1}{2}\cdot2^n-2$.

\begin{figure}[H]
    \centering
    \leavevmode
\annotate{
\Qcircuit @C=0.5em @R=.7em {
&\qw &\multigate{1}{U}   & \qw   & 
& &\ctrlb{1}        &\gate{\tilde{U}} & \ctrlb{1}        & \ctrlb{1} &\gate{R_z} & \qw\\
&\qw   &\ghost{U}          &\qw    &\push{\rule{.1em}{0em}\raisebox{2.2em}{=}\rule{.1em}{0em}}
& &\gate{\tilde{U}} &\ctrlb{-1}      &\gate{\tilde{U}}  &\gate{R_z} & \qw & \qw }
}{A circuit equivalence is shown, with on the left side of the equal sign a 2-qubit gate, which is called U. On the right of the equal sign is a quantum circuit with 2 qubits. This circuit consists of the following gates, from left to right: three 1-qubit tilde(U) gates on the bottom, the top and then the bottom qubit. They are each uniformly controlled by the other qubit. Then an R\_z gate on the bottom qubit, uniformly controlled by the top qubit and a regular R\_z gate on the top qubit.}
    \caption{Recursive CSD decomposition \cite{art:decompositionofgeneralquantumgates}}
   \label{circ:recurviveCSD}
\end{figure}
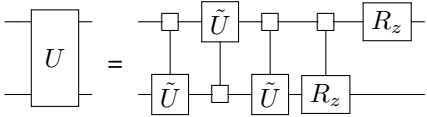

\subsection{Quantum Shannon Decomposition} \label{subsec:qsd} 
\begin{figure}[H]
    \centering
    \leavevmode
\annotate{
\Qcircuit @C=0.5em @R=.7em {
&\qw &\multigate{1}{U}   & \qw   & 
& &\qw &\qw & \gate{R_z} &\qw & \gate{R_y} &\qw & \gate{R_z} & \qw &\qw\\
&\qw   &\ghost{U}          &\qw    &\push{\rule{.1em}{0em}\raisebox{2.2em}{=}\rule{.1em}{0em}}
& &{}\qw &\gate{G_1} & \ctrlb{-1} &\gate{G_2} & \ctrlb{-1} &\gate{G_3} & \ctrlb{-1} & \gate{G_4}&\qw }
}{A circuit equivalence is shown, with on the left side of the equal sign a 2-qubit gate, which is called U. On the right of the equal sign is a quantum circuit with 2 qubits. This circuit consists of the following gates: a 4 1-qubit gates called G1, G2, G3 and G4 on the bottom qubit. Between these are a uniformly controlled rotation gates on the top qubit controlled by the bottom qubit, in order an R\_z, and R\_y and an R\_z gate.}
    \caption{Quantum Shannon Decomposition\cite{synthesisofquantumlogiccircuits}}
    \label{fig:quantumshannon}
\end{figure}
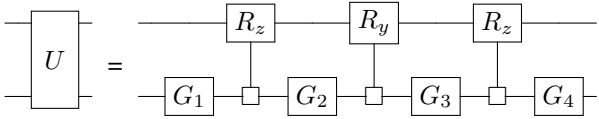

\cite{synthesisofquantumlogiccircuits} introduces another way of using the CSD from \cref{subsec:csdandnq}, called Quantum Shannon Decomposition (QSD). The decomposition of a 2-qubit gate is shown in \cref{fig:quantumshannon}. 

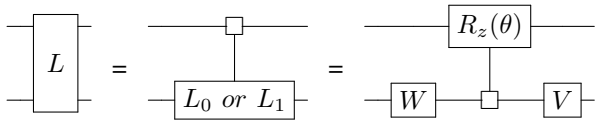
\begin{figure}[H]
    \centering
    \leavevmode
\annotate{
\Qcircuit @C=0.5em @R=.7em {
&\qw &\multigate{1}{L}   &\qw    &
& &\qw &\ctrlb{1}    &\qw        &
& & \qw &\qw        & \gate{R_z(\theta)}              & \qw & \qw\\
&\qw  &\ghost{L}          &\qw    &\push{\rule{.3em}{0em}\raisebox{2.2em}{=}\rule{.3em}{0em}} 
& &\qw & \gate{L_0~or~L_1}  &\qw        &\push{\rule{.3em}{0em}\raisebox{2.2em}{=}\rule{.3em}{0em}}
& &\qw &\gate{W}   & \ctrlb{-1} & \gate{V} &\qw
}
}{A circuit equivalence is shown, with on the left side of the equal sign a 2-qubit gate, which is called L. On the right of the equal sign is a quantum circuit with 2 qubits. It shows a uniformly controlled gate on the bottom qubit, labelled "L0 or L1", controlled by the top qubit. On the right of this is a second equal sign. It shows that these circuits are equivalent to a quantum circuit with two qubits, with first a 1-qubit gate called W on the bottom qubit, then a multi-controlled R\_y(theta) gate on the top qubit, controlled by the bottom one, and a 1-qubit called V on the bottom qubit.}
\caption{Decomposition of the $L$ matrix in QSD \cite{synthesisofquantumlogiccircuits}.}
\label{fig:decompofL}
\end{figure}

The start of the decomposition is the same as in \cref{subsec:csdandnq}, but the $L$ and $R$ matrices are decomposed using Eigenvalue decomposition. This is shown in \cref{fig:decompofL}. The resulting matrices are unitary gates applied to one less qubit than the starting unitary. This leads to the circuit in \cref{fig:quantumshannon}, where the $D$-matrix is implemented as a multi-controlled $R_z$ gate. 

Quantum Shannon Decomposition is applied recursively until the final 1-qubit gates can be implemented with ZYZ decomposition. This means only the multi-controlled rotation gates contribute to the number of CNOTs, each of which requires $2^{n-1}$ CNOT gates for a single step of the recursion of an n-qubit gate. This leads to a total CNOT count of $\sfrac{3}{4}\cdot 4^n-\sfrac{3}{2}\cdot 2^n$ for this decomposition method.

There are two optimizations that can be implemented on top of this implementation of Quantum Shannon Decomposition. The first is to stop the recursion at 2-qubit gates, and translate those as in \cref{subsec:decompoftwoqubitgates}.
The second optimization is to implement the central multi-controlled $R_z$ gate using CZ gates rather than CNOTs, of which one can be absorbed into the neighboring multiplexer. This results in one less CNOT gate at each level of the recursion. With these two implementations the CNOT count comes to $\sfrac{23}{48} \cdot 4^n - \sfrac{3}{2} \cdot 2^n + \sfrac{4}{3}$\cite{synthesisofquantumlogiccircuits} .



\begin{table*}[!htb]
\centering
\caption{CNOT counts for different implementations of unitary decomposition for a 1 through 5-qubit, as well as an n-qubit unitary gate.}
\label{tab:cnotcounttotal}
\begin{tabular}{lrrrrrlr}
\toprule
Number of qubits                               & 1          & 2          & 3           & 4            & 5             & $n$ & Section\\
\midrule
Theoretical lower bounds  \cite{art:minimaluniversal}                               & 0          & 3          & 14          & 61           & 252       & $\tfrac{1}{4}\cdot(4^n-3n-1)$    &\ref{subsec:theoreticallowerbound}-\ref{subsec:decompoftwoqubitgates}   \\ 
Iterative unentangling \cite{synthesisofquantumlogiccircuits}
                                   & 0          & 8          & 62          & 344          & 1642     &$2\cdot4^n-(2n+3)$$\cdot$$2^n+2n$   &\ref{subsec:theorem9}    \\ 
Givens rotations \cite{art:efficientdecompositionofquantumgates}   & 0          & 4          & 64          & 536          & 4156     &$\approx 8.4\cdot 4^n$    &\ref{subsec:effdecomp} \\ 
Recursive CSD \cite{misc:rudimentaryquantumcompiler}     & 0        & 14          & 92          & 504          & 2544        & $\tfrac{1}{2}\cdot n\cdot4^n-\tfrac{1}{2}\cdot2^n$     &\ref{subsec:csdandnq} \\ 
Recursive CSD (optimized) \cite{art:decompositionofgeneralquantumgates}     & 0        & 4          & 26          & 118          & 494        & $\tfrac{1}{2}\cdot4^n-\tfrac{1}{2}\cdot2^n -2$     &\ref{subsec:csdandnq} \\ 
\textbf{QSD} \cite{synthesisofquantumlogiccircuits}                           & \textbf{0} & \textbf{6} & \textbf{36} & \textbf{168} & \textbf{720}  & $\tfrac{3}{4}\cdot 4^n -\tfrac{3}{2}\cdot 2^n$ & \ref{subsec:qsd} \\ 
QSD (optimized) \cite{synthesisofquantumlogiccircuits}                               & 0          & 3          & 20          & 100          & 444    & $\tfrac{23}{48}\cdot 4^n-\tfrac{3}{2}\cdot 2^n + \tfrac{4}{3}  $ & \ref{subsec:qsd}    \\ \bottomrule
\end{tabular}%
\end{table*}%

\begin{figure}[!htb]
    \centering
\annotate{
    \includegraphics[width=\linewidth]{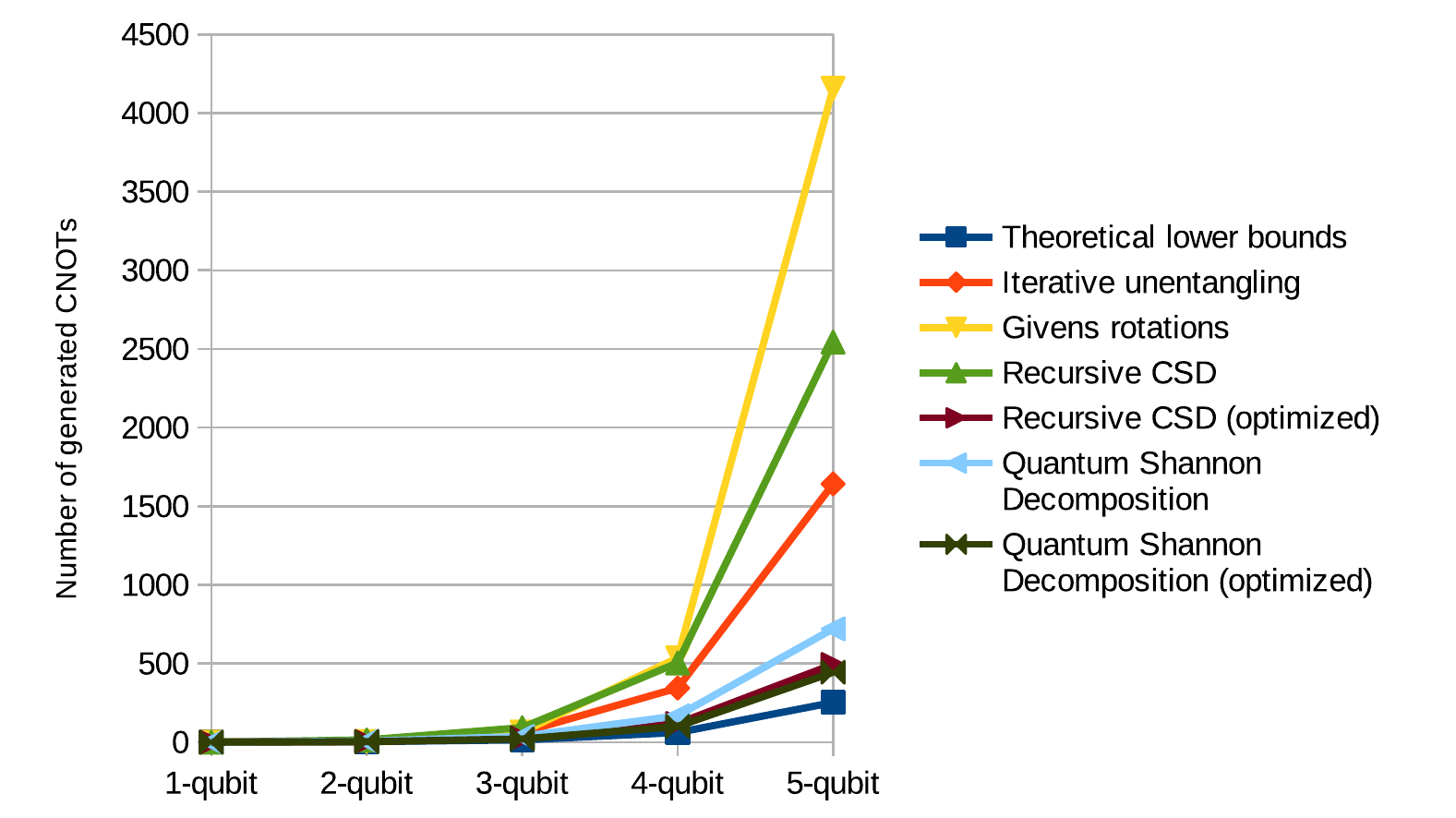}
}{The number of CNOTs for each implementation as in the table, plotted in a line chart. The x-axis is labelled from 1-qubit to 5-qubit, and the y-axis is labelled "Number of generated CNOTs", and goes from 0 to 4500. The plot shows that all methods for unitary decomposition generate an exponential number of CNOT gates with respect to the number of qubits. The most CNOTs are generated when using Givens rotations, and the lowest are the theoretical lower bounds.}
    \caption{CNOT counts for different implementations of unitary decomposition for 1 through 5-qubit gates}
    \label{fig:gatecounts}
\end{figure}

\subsection{Selection of the algorithm} \label{subsec:selection}
For each decomposition method, the CNOT gate counts are compiled in \cref{tab:cnotcounttotal} and plotted in \cref{fig:gatecounts}. As an indication, the number of CNOT gates resulting from the decomposition of a 1 to 5-qubit unitary gate is given. Along with the general formulas for the number of CNOT gates resulting from the decomposition of an n-qubit gate, if such a formula was available. %

As can be seen in \cref{tab:cnotcounttotal}, the optimized version of QSD results in the fewest CNOT gates. The choice was therefore made to implement this decomposition, although not the optimized version. The optimizations from \cite{art:efficientdecompositionofquantumgates} can be implemented without any modifications to a base implementation of the algorithm.

Besides that, QSD has several other advantages. The recursion is performed at general n-qubit gates rather than multi-controlled 1-qubit gates, which makes it relatively simple to implement. If algorithmic implementations for 3-qubit, 4-qubit or 5-qubit or bigger general gates are found, they can be easily implemented. The same goes for other specific optimizations. And because the mathematical decompositions are done separately for each step in the recursion, rather than all at once at the beginning, any underlying structure in the beginning or intermediate matrices can be taken advantage of immediately, therefore potentially skipping many computational steps as well as decreasing the size of the resulting circuit. 

For these reasons, the choice was made to go with Quantum Shannon Decomposition for the implementation of unitary decomposition in OpenQL.%

\lstMakeShortInline[basicstyle=\normalsize\ttfamily,language=none]|
\section{Implementation} \label{sec:implementation}
The implementation of the decomposition in OpenQL is split into two parts: the calculation of all of the rotation angles, and the generation of the circuit. This is done so that the implementation is independent from OpenQL. 
A short example of unitary decomposition in OpenQL is shown in Code Example \ref{lst:unidecompopenql}, the full code can be found in Code Example \ref{lst:timingprogram}.

\begin{lstlisting}[language=Python,morekeywords={Unitary, decompose, add_kernel}, caption={Using unitary decomposition in OpenQL.},label={lst:unidecompopenql}]
  import openql.openql as ql
  # OpenQL preamble
  unitary = ql.Unitary('u1',[0.584-0.387j, -0.371-0.610j, 
                            -0.254-0.667j,  0.664-0.223j])
  unitary.decompose()
  kernel.gate(unitary, [0])

  program.add_kernel(kernel)
  compiler.compile(program)
\end{lstlisting}%

For unitary decomposition in OpenQl, first a \textit{Unitary} object is defined, which is then decomposed to calculate all the angles for all the rotation gates. The \textit{Unitary} is then added to a \textit{kernel} as any other gate. The \textit{kernel} is added to a \textit{program}, which is compiled with a \textit{compiler}. 
The implementation is thus split between the \textit{Unitary} class and the call to \textit{kernel.gate()}.

\subsection{The Unitary class} \label{subsec:unitaryopenql}
The \textit{Unitary} is instantiated with a string and an array. The content of this array is the unitary matrix, which is of size $2^n\times2^n$ for an n-qubit gate. The complete Quantum Shannon Decomposition is computed only when ``decompose()'' is called, and the calculated angles for the resulting rotation gates are added to a list. This is done so that the \textit{Unitary} can be used multiple times in a program without recalculation of the whole decomposition.

But before the decomposition is started, it is first checked if the input matrix is unitary. If this is the case, all of the intermediate matrices will also be unitary \cite{art:historyandgeneralityofthecsdecomposition}, so this check is only necessary once. Furthermore, all of the $M^k$ (Gray code) matrices, which are needed for the multi-controlled rotation gates, are added to a lookup table so they do not need to be calculated anew at each decomposition step. 

To make certain that the decomposition is correct, each single intermediate decomposition is checked. For each step only three matrices need to be multiplied, and this saves any calculations that might be done on an incorrect matrix. If any step of the decomposition is not correct, an exception is thrown and the decomposition is stopped. 

The Eigen\cite{misc:eigendoc} library is used to do Singular Value Decomposition (SVD), eigenvalue decomposition and matrix multiplication. The recursion is centered on a main function, which takes as parameters a unitary matrix and the number of qubits. The latter is to keep track of the level of recursion. 


Computation of the CSD is done using the method from \cite{art:historyandgeneralityofthecsdecomposition}, which uses SVD. The demultiplexing function uses Schur matrix decomposition for the (sub)matrices smaller than $2^6 \times 2^6$, and eigenvalue decomposition for bigger matrices. This is done because Schur matrix decomposition is faster for small matrices \cite{misc:eigendoc}.

The algorithm is recursive, and the demultiplexing step calls on the main function again for the decomposition of the smaller unitary matrices. If the matrices are of size $2\times2$, the rotation angle for the 1-qubit rotation gates are calculated using ZYZ decomposition as in \cref{subsec:zyz}.

Because the \textit{Unitary} does not have access to the qubit numbers of the circuit, only the angles for the multi-controlled $R_y$ and $R_z$ are calculated at this point. This is done as in \cref{subsec:multicontrolledgateS}, through solving the following matrix equalities:

\begin{align}
    M^k \begin{bmatrix}\theta_1 \\ \vdots \\\theta_{2^k} \end{bmatrix} = \begin{bmatrix}\alpha_1 \\ \vdots \\ \alpha_{2^k} \end{bmatrix} \label{eq:Mk}
\end{align}
where $M^k$ is a square matrix where all the entries are either "+1" or "-1", which are calculated using Gray code using \cref{eq:Mk2}. 
%
\begin{align}
    M^k_{ij} = (-1)^{b_(i-1)\cdot \gamma_(j-1)} \label{eq:Mk2}
\end{align}
where the exponent is the bit-wise inner product of two binary vectors: $b_i$ and $\gamma_j$. $b_i$ is the integer $i$ and $\gamma_j$ is the $j$th value of the Gray code.

For the multi-controlled $R_y$ gate the values of $\alpha_i$ are calculated by taking the arc sine of the diagonal entries of the $S$-matrix from the CSD.%
\begin{align}
    \begin{bmatrix}\alpha_1 \\ \vdots\\ \alpha_{2^k} \end{bmatrix} = 2\cdot arcsin\left(S_{i,i}\right) \label{eq:multicontrolledY}
\end{align}
For the multi-controlled $R_z$ gates the values of $\alpha_i$ is calculated by taking the natural logarithm of the $D$-matrix from the demultiplexing.  
\begin{align}
    \begin{bmatrix}\alpha_1 \\ \vdots\\ \alpha_{2^k} \end{bmatrix} = -2\cdot\sqrt{\text{-}1}\cdot ln\left(D_{(i,i)}\right) \label{eq:multicontrolledZ}
\end{align}
All the angles for all rotation gates are added to a list, which is used to generate the correct gates when the \textit{Unitary} is added to a circuit.

\subsection{kernel.gate()}
At the \textit{kernel} level, when the (decomposed) \textit{Unitary} object is added to the circuit, the gates and CNOTs are assembled and added to the circuit list. 
At this point, it is checked whether the \textit{Unitary} is decomposed and if it is applied to the correct number of qubits.  The first is checked from a flag that is set to ``true" at the end of the decomposition. The latter is calculated from the size of the unitary matrix, which should be $2^n \times 2^n$ for an n-qubit gate.

Because the \textit{kernel} only has the qubit numbers and the list of rotation angles, it does not have insight into whether any optimizations have happened. Therefore, the gates are added purely sequentially to the circuit, and each recursive call to the main function returns the total number of rotation angles that was used up until that point. If gates have been removed by an optimization, a specific angle is added to the circuit which signals how many gates have been removed, and these gates are skipped during circuit generation.

It is expected that the decomposition will take the most time to compute, as well as the most memory, since it contains the mathematical algorithms and matrix multiplications. Comparatively, using the calculated angles to make the circuit will not require much time or memory. So adding the circuit sequentially is not expected to have much of an impact on the total resources required by the circuit, while it allows for a much more modular implementation of unitary decomposition.

        

\subsection{Compilation of the OpenQL program}
After all gates have been added to the circuit, the \textit{kernel} is added to a \textit{program} which is compiled in OpenQL. From this point, the gates from the decomposition are handled in the same way as any manually added gates. So the features and optimizations from the lower levels of the programming language can be fully used for the circuit\cite{art:openql2020}. Afterwards, the circuit is transformed into quantum assembly language and written to an output file as usual, or directly passed on to the simulator. 

\section{Implementation optimization} \label{sec:optimization}

For execution of the resulting circuit, it is important that it is as short as possible for the reasons mentioned in \cref{sec:comparison}. To this end, the algorithm itself was selected to generate as few gates as possible. Combining and removing  individual gates will be done at a later compile step by the OpenQL compiler \cite{openql:2020}, but more structural optimizations can be done during the decomposition. For example, QAM, one of the algorithms from \cref{sec:whyunittdecomp}, generates a unitary matrix that has an internal structure which can be used to skip many steps in the recursion (see \cref{sec:whyunittdecomp}). The implemented optimizations take advantage of the matrix structure through early detection of multiplexers and the detection of unaffected qubits. 

\subsection{Detection of multiplexers}
Before the CSD is started, it is checked whether the upper right and lower left quarters of the matrix are already zero-matrices.
If that is the case, the matrix already has the structure of a multiplexer, and is directly passed to the demultiplexing step. This is signaled to the kernel by adding a specific gate angle to the list of rotation angles. This operation halves the number of resulting gates for this step of the decomposition. 

\subsection{Unaffected qubits}
If a decomposition step leaves a qubit unaffected, then it is not necessary to apply any gates to that qubit, and an n-qubit gate can be handled as an (n-1)-qubit gate. This reduces the resulting number of gates for this step by more than $\sfrac{3}{4}$. So before the main decomposition is called, it is checked if the matrix is of the form $A \oplus I$ or $I \oplus A$. Each step of the QSD evaluates unitary gates on one less qubit, so any unaffected qubits become the first or last qubit at some point in the decomposition. If an unaffected qubit is detected, this is also signaled to the kernel. 
The unitary matrix of size (n-1) is then assembled and passed back to the main function of the decomposition. 

\subsection{Execution time optimizations}
There are also some optimizations to reduce the execution time and memory use of the decomposition. 

One of the things done to reduce the total execution time and memory use is the fitting of ``.noalias()'' flags to all places where the product of multiple matrices is assigned to a different matrix. The Eigen library assumes aliasing for all such operations and without this flag, it evaluates the result of a matrix product into a temporary matrix that is then copied\cite{misc:eigendoc}. Another optimization is that all matrices are passed as references where possible, to prevent any unnecessary copying of data. 

The execution time and memory use of the decomposition after these and other optimizations can be found in \cref{sec:results}. For most of the decomposition the total execution time scales with approximately $4^n$ for an $2^n \times 2^n$ unitary matrix, which corresponds to an n-qubit gate. This is a linear relation with the number of generated gates and the number of elements in the input matrix.

\begin{figure*}[ht]
    \centering
    \annotate{
    \resizebox{\textwidth}{!}{%
    \includegraphics[height=3cm]{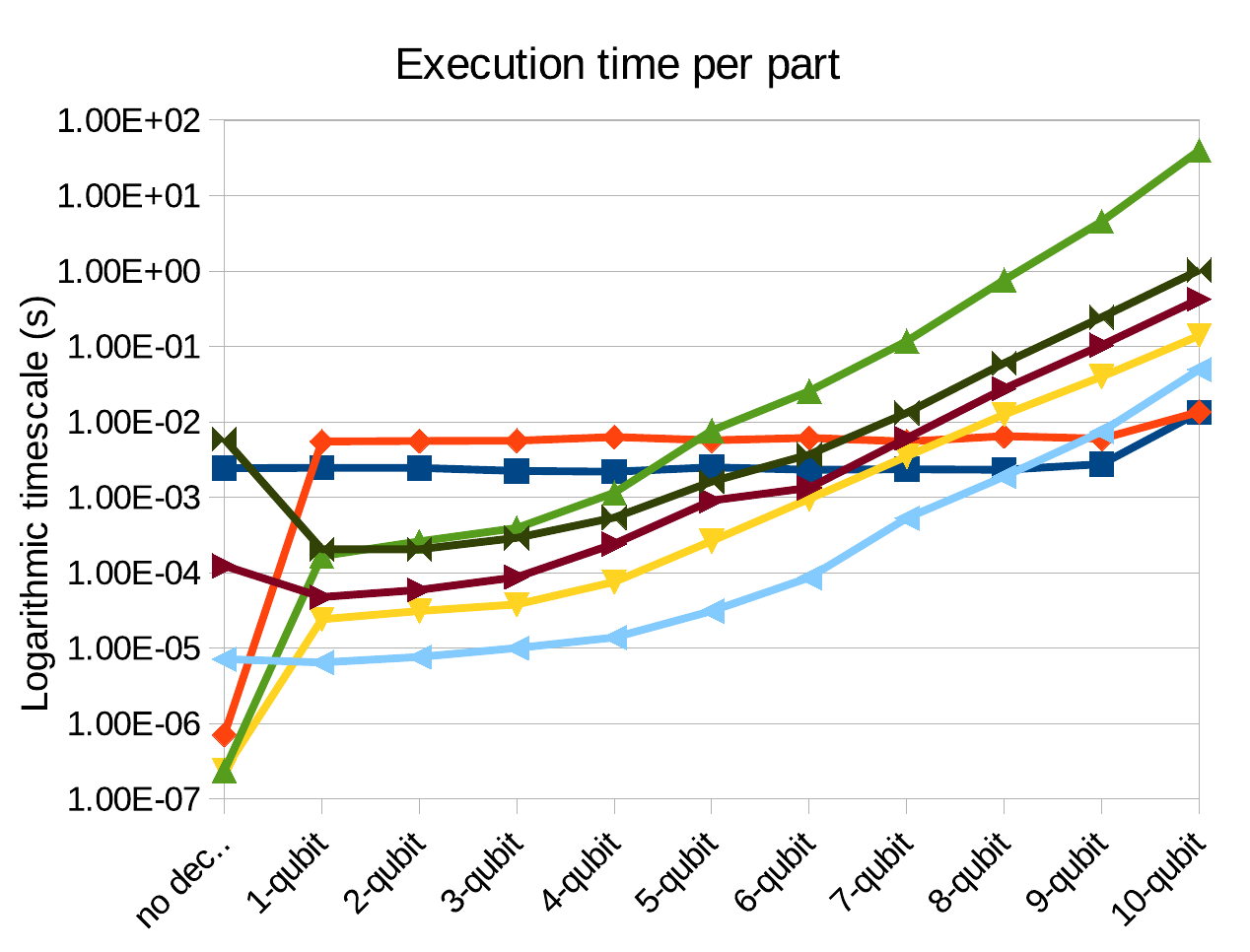}%
\quad
    \includegraphics[height=3cm]{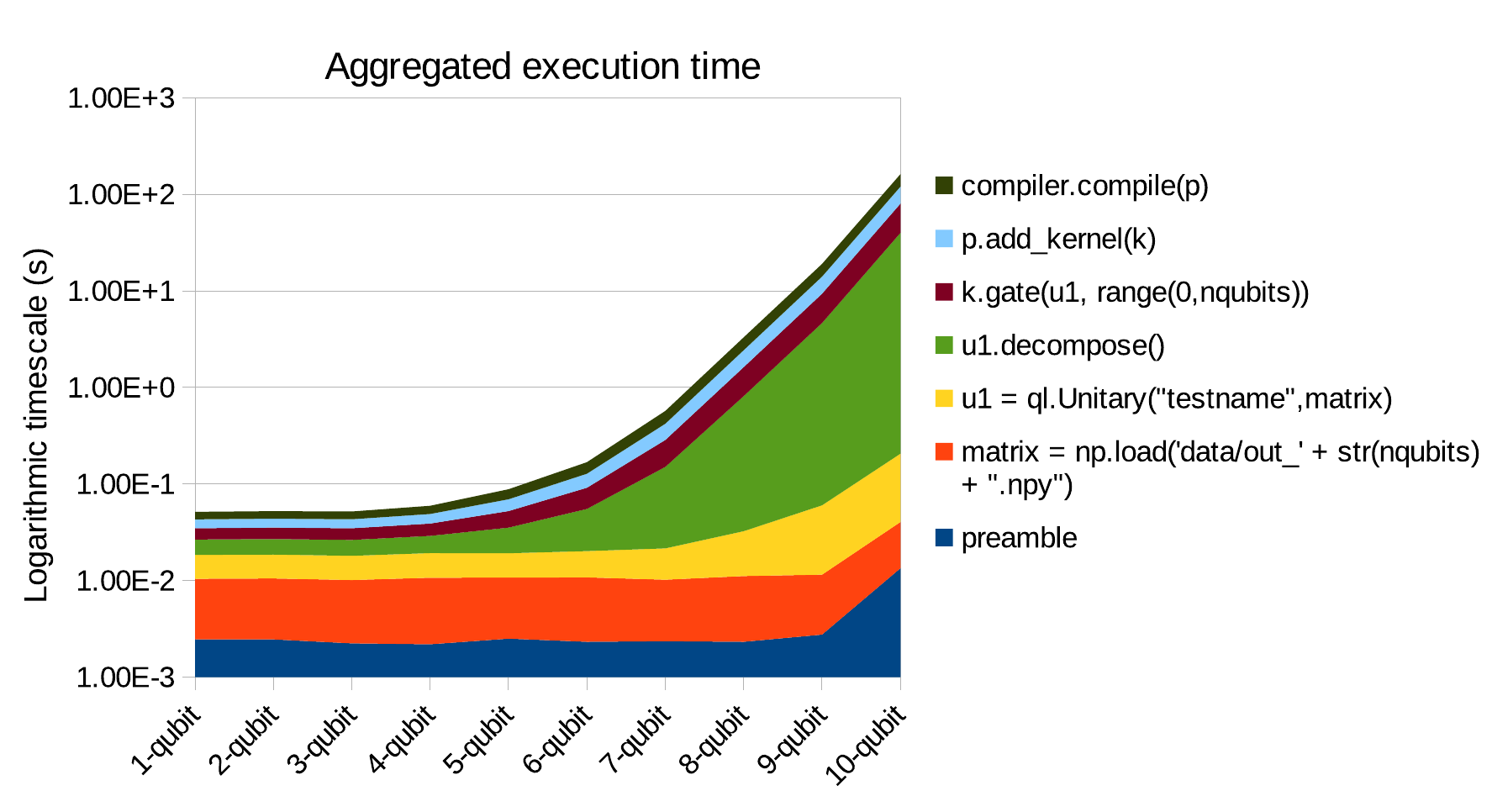}%
}
}{Two plots are shown side by side. The left one is labeled "Execution time per part", and the right "Aggregated execution time". Both plot on the x-axis the size of the unitary matrix from 1-qubit to 10-qubits,a and on the y-axis the logarithmic execution time in seconds. They show that the main part of the execution time is used by the "u1.decompose()" line. The other parts of the implementation also take longer for bigger sizes of input matrices, at a similar rate as the decompose step.}
    \caption{Execution time for the timed intervals, for different sizes of unitary matrices}
    \label{fig:executiontimes}
\end{figure*}



\section{Results} \label{sec:results}
The execution time of different parts of the decomposition is measured as the elapsed wall-clock time, with measurements in between function calls to determine the relative time consumption. The final execution times are shown in \cref{fig:executiontimes}. These tests were executed using a Dell Latitude 7400 with an 8th Generation Intel® Core™ i7-8665U Processor and 2x 4GiB DDR4 RAM. 

The program in Code Example \ref{lst:timingprogram} has been used to determine the execution time and memory used by the decomposition. To measure execution time, the Python "time" package is used to determine the time difference between the start and various points of the program. The time for each part of the code, as well as the resulting aggregated execution time, can be found in \cref{fig:executiontimes} and \cref{tabexecutiontimes}. 

As expected, the decomposition itself takes the most time, more than 10 times that of any other part. This is because of the considerable mathematical decompositions and the number of matrix operations. One of the algorithms used in the decomposition is eigenvalue decomposition, which is an iterative algorithm that requires $O(6^n)$ operations for an $2^n \times 2^n$ matrix \cite{misc:lapackusersguide}. 
The data also shows that generation of the rotation gates and CNOTs does not contribute much to the total execution time of the algorithm, as expected.
And since the complete decomposition is calculated at design time, it does not influence the run-time of the final circuit when it is executed on a quantum accelerator. 
\begin{figure}[ht]
    \centering
\annotate{
\includegraphics[width=\linewidth]{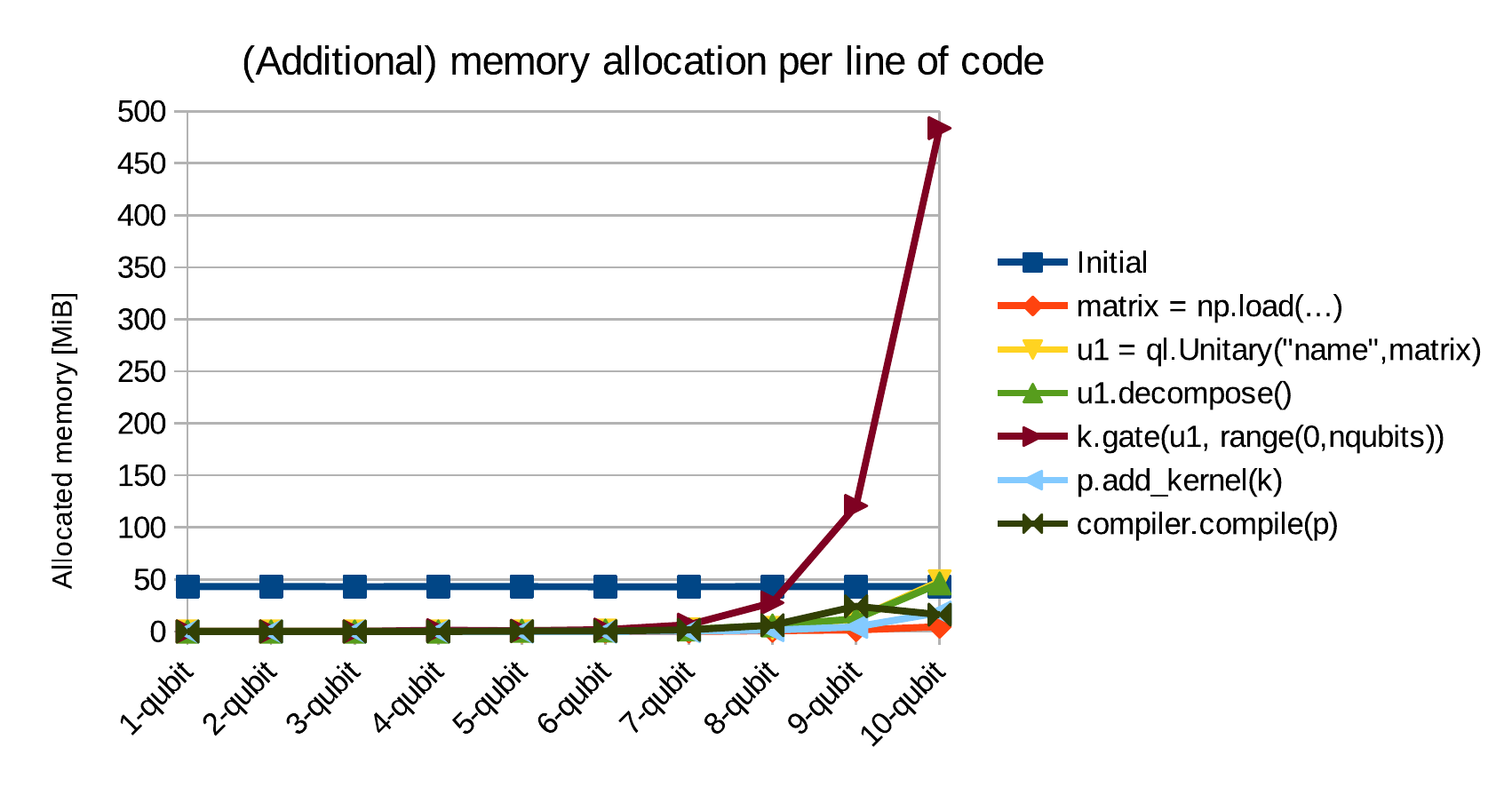}
}{A line plot is shown. On the x-axis is again 1 through 10-qubits. The y-axis is labeled "Allocated memory [MiB]" and goes from 0 to 500. Each line in the plot corresponds to the additional memory allocated for a line in the test program. The plot shows an exponential relation for the "k.gate" line, which climbs to almost 500 MiB, which the rest of the lines do not exceed 40 MiB.}
\caption{Additional memory allocated per line, for different sizes of unitary matrices}
\label{fig:memoryallocation}
\end{figure}%
The same program has also been used to determine the memory allocation. This has been measured using the Python memory\_profiler package. The results of this are shown in \cref{tab:memoryallocation} and \cref{fig:memoryallocation}. After an initial allocation of about 40 MiB, noteworthy additional allocation of memory occurs only when k.gate(...) is called. 
This means that the complete unitary decomposition requires much less memory than generating and storing the resulting circuit in OpenQL. 

\section{Comparison to other implementations} \label{sec:comparequbiter}
We also compared our implementation of unitary decomposition to that of Qubiter \cite{misc:rudimentaryquantumcompiler}. Qubiter is a quantum compiler/programming language that aims to provide a set of tools for designing and simulating quantum circuits. As part of that, they offer unitary decomposition based on the recursive CSD from \cref{subsec:csdandnq}. It is the only other programming language that offers a decomposition implementation focused on generated circuit length. Qubiter is written in Python and uses numpy for the mathematics, as well as the LAPACK cuncsd function for the CSD\cite{misc:qubitergithub}.

\begin{figure}[ht]
    \centering
    \annotate{
    \includegraphics[width=0\linewidth]{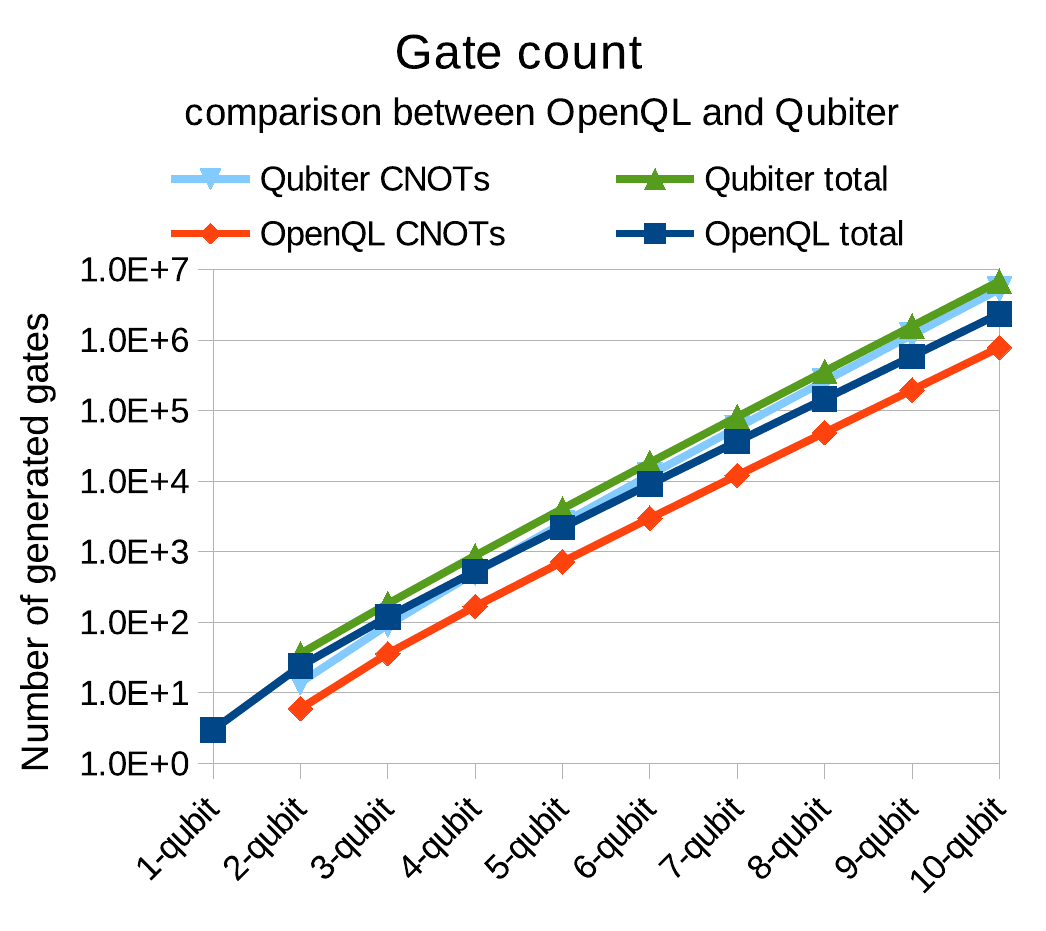}
    }{A line plot is shown, with as title: Gate count comparison between OpenQL and Qubiter. On the x-axis is again 1 through 10-qubits. On the y-axis are the number of generated gates on a logarithmic scale, from 1.0E+0 to 1.0E+7. The plot shows 4 almost straight lines, that are sloping a bit downward for bigger sizes of unitary matrices. The highest line, which corresponds to the most generated gates, is Qubiter total. A small fraction below it is Qubiter CNOTs. Below that is OpenQL total and the lowest line is OpenQL CNOTs.}
    \caption{Number of generated CNOTs and total gates for OpenQL and Qubiter from the decomposition of different sizes of unitary matrices.}
    \label{fig:openqlqubitergatecounts}
\end{figure}

Because we use QSD in our implementation of unitary decomposition in OpenQL, the decomposition generates much shorter circuits than the one in Qubiter.
To get the total gate count for both languages, the number of lines in the output text files have been counted. Both can generate a file with a representation of the quantum circuit, with each gate on a separate line. The total gate count also includes rotation gates and not just CNOTs. The results for OpenQL and Qubiter are plotted in \cref{fig:openqlqubitergatecounts} and can be found in \cref{tab:openqlqubitergates}. It is clear that OpenQL always generates fewer gates than Qubiter, and almost all of the difference is in the number of CNOTs. For a 10-qubit gate, unitary decomposition with OpenQL generates half as many CNOTs as Qubiter, and produces a total circuit that is almost 3 times as short.

\begin{figure}[ht]
    \centering
\annotate{
    \includegraphics[width=\linewidth]{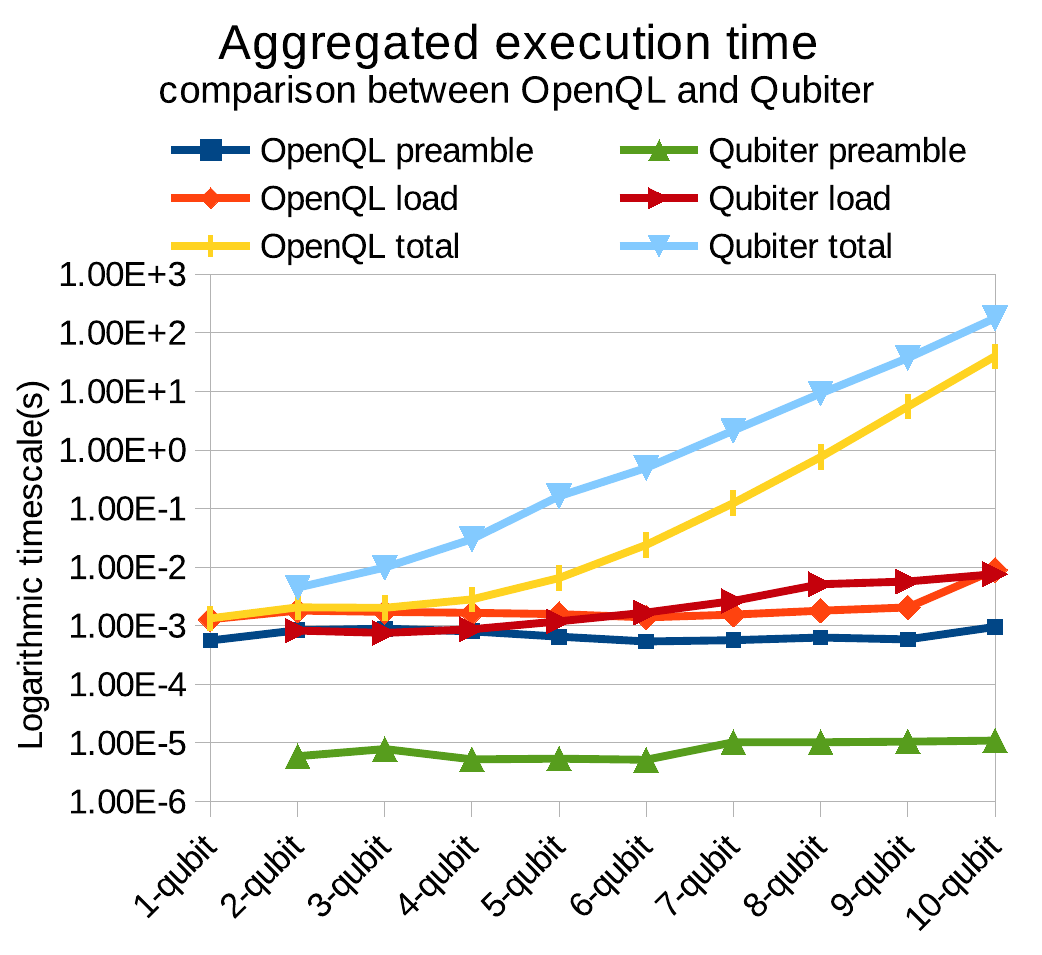}
}{A line plot is shown, with as title: Aggregated execution time comparison between OpenQL and Qubiter. On the x-axis is again 1 through 10-qubits. The y-axis is labeled "Logarithmic timescale(s)" and it goes from 1.0E-6 to 1.0E3. The plot shows 6 lines, for the aggregated execution times for the preamble, load and total time required by the decomposition for both OpenQL and Qubiter. The preamble and load lines are all approximately flat an do not exceed 2E-2 seconds. The total time for OpenQL is line that is slightly sloping upwards, that starts at 1E-3 that goes to 40 seconds at the 10-qubit mark. The line for qubiter total is a straight line that goes from 5E-3 to 180 seconds.}
    \caption{Execution time of the preamble, matrix load times and total decomposition for OpenQL and Qubiter for different sizes of unitary matrices.}
    \label{fig:openqlqubitercomparison}
\end{figure}

The implementations are also compared on the time used to compute the unitary decompositions. The aggregated execution times for decompositions of 2 to 10-qubit unitary gates can be found in \cref{tab:openqlqubitercomparison}, and are plotted in \cref{fig:openqlqubitercomparison}. The execution time of both decompositions scale approximately linearly with the input matrix size. 
As can be seen in the table and the figure, OpenQL is considerably faster than Qubiter. When comparing the total execution times, it becomes clear that the OpenQL implementation takes more time per input matrix element ($4^n$) due to the Eigenvalue decomposition. Qubiter does not have that issue, but using unitary decomposition in OpenQL is about 10 to a 100 times faster for the decomposition of 1 to 10-qubit unitary gates. In addition to being faster, unitary decomposition in OpenQL generates a much shorter circuit for all sizes of unitary matrices.
\section{Conclusion and future work} \label{sec:conclusion}
With the implementation of unitary decomposition, OpenQL can be used for any quantum algorithm that uses arbitrary unitary gates. One such algorithm is QiBAM \cite{art:qibam}, which can now be implemented using OpenQL. 
This is not possible without unitary decomposition. The decomposition generates more gates than the theoretical minimum, but the structure of the decomposition means that further optimizations can be easily integrated with the current implementation. The decomposition is done using Quantum Shannon Decomposition, which is  up to $10\times$ more efficient in number of generated gates than other examined algorithms. Two optimizations were implemented to take advantage of the internal structure of the input or intermediate unitary matrices, which can drastically reduce the length of the resulting circuit. With these optimizations, the final resulting gate count can be much lower than the illustrated worst case numbers. 

The decomposition results in $O(\tfrac{3}{4}4^n)$ CNOT gates and $O(\tfrac{9}{4}4^n)$ total gates. Compared to other implementations of unitary decomposition, specifically Qubiter, it generates half the number of CNOTs and a total circuit that is three times as short for the decomposition of 10-qubit gates. Although the execution time of the decomposition scales with $O(6^n)$ for matrices of size $2^n\times 2^n$, for the decomposition of up to 10-qubit gates our implementation is 10-100 times faster than Qubiter.

There are several avenues that can further bring down the number of gates the decomposition generates, which are:
\begin{itemize}
    \item Implementation of a minimum 2-qubit circuit, such as the one described in  \cite{synthesisofquantumlogiccircuits}.
    \item Additionally, implementation of a universal 3-qubit gate, such as the one in \cite{art:realizationofathreequbitgate}.
    \item Implementing the multiplexed $R_z$ gate with a CZ gate, as expressed in \cite{synthesisofquantumlogiccircuits}.
    \item Reworking the QSD so that the intermediate matrices cancel out, as the input matrix has fewer degrees of freedom than the matrices resulting from the QSD. Therefore, it might be possible to choose some of these intermediate matrices in such a way that they can be decomposed using fewer elementary gates. 
    \item Implementation of other specific efficient decompositions, such as controlled unitary gates (as opposed to uniformly controlled gates), quantum multiplexers or specialized multi-controlled rotation gates.
\end{itemize}
Another possibility to bring down gate count is to implement other application specific optimizations when the input for the unitary decomposition is known to have more constraints. Such as arbitrary unitary gates that only apply right-angle operations to the qubits, or matrices that are Hermitian\footnote{A Hermitian matrix is a square, complex matrix ($H$) that is equal to its conjugate transpose ($H^\dagger$), i.e. $H=H^\dagger$ \cite{book:gelfand1989lectures}} as well as unitary. The decomposition might be tailored to take advantage of these constraints, so that the decomposition of these more constrained input matrices results in shorter circuits than the implemented general unitary decomposition.

For near-term quantum applications, the decomposition generates too many gates for unitary matrices bigger than a certain size. Although the precise limit depends on the specific implementation, the decomposition of a 3-qubit unitary gate might already result in a circuit that is too long. But there are several optimizations that can be done to make unitary decomposition more feasible for near-term quantum applications with non-perfect qubits, such as modifying the decomposition to generate a more parallel circuit, or splitting the resulting circuit in several pieces that can be executed separately. Nearest-neighbor circuits can be used to minimize the cost of mapping.
And due to the identical structure for each decomposed circuit, the structure of a real quantum system can be adjusted so that it perfectly fits unitary decomposition, which can reduce or completely remove the need for mapping operations.
 
Ultimately, the goal of all of these suggestions is to keep unitary decomposition and OpenQL relevant and useful both for near-term and future quantum applications.

\bibliographystyle{IEEEtran}
\bibliography{references.bib}

\begin{thebibliography}{10}
\providecommand{\url}[1]{#1}
\csname url@samestyle\endcsname
\providecommand{\newblock}{\relax}
\providecommand{\bibinfo}[2]{#2}
\providecommand{\BIBentrySTDinterwordspacing}{\spaceskip=0pt\relax}
\providecommand{\BIBentryALTinterwordstretchfactor}{4}
\providecommand{\BIBentryALTinterwordspacing}{\spaceskip=\fontdimen2\font plus
\BIBentryALTinterwordstretchfactor\fontdimen3\font minus
  \fontdimen4\font\relax}
\providecommand{\BIBforeignlanguage}[2]{{%
\expandafter\ifx\csname l@#1\endcsname\relax
\typeout{** WARNING: IEEEtran.bst: No hyphenation pattern has been}%
\typeout{** loaded for the language `#1'. Using the pattern for}%
\typeout{** the default language instead.}%
\else
\language=\csname l@#1\endcsname
\fi
#2}}
\providecommand{\BIBdecl}{\relax}
\BIBdecl

\bibitem{Shor:1997}
\BIBentryALTinterwordspacing
P.~W. Shor, ``Polynomial-time algorithms for prime factorization and discrete
  logarithms on a quantum computer,'' \emph{SIAM J. Comput.}, vol.~26, no.~5,
  pp. 1484--1509, Oct. 1997. [Online]. Available:
  \url{http://dx.doi.org/10.1137/S0097539795293172}
\BIBentrySTDinterwordspacing

\bibitem{grover:1997}
\BIBentryALTinterwordspacing
L.~K. Grover, ``{Quantum mechanics helps in searching for a needle in a
  haystack},'' \emph{Physical Review Letters}, vol.~79, no.~2, pp. 325--328,
  1997. [Online]. Available:
  \url{http://link.aps.org/doi/10.1103/PhysRevLett.79.325}
\BIBentrySTDinterwordspacing

\bibitem{art:realizingquantumalgorithms}
C.~G. {Almudever}, L.~{Lao}, R.~{Wille}, and G.~G. {Guerreschi}, ``Realizing
  quantum algorithms on real quantum computing devices,'' in \emph{2020 Design,
  Automation Test in Europe Conference Exhibition (DATE)}, 2020, pp. 864--872.

\bibitem{art:notallqubitsequal}
S.~Tannu and M.~Qureshi, ``Not all qubits are created equal: A case for
  variability-aware policies for nisq-era quantum computers,'' 04 2019, pp.
  987--999.

\bibitem{openql:2020}
N.~Khammassi, I.~Ashraf, J.~v.~Someren, R.~Nane, A.~M. Krol, M.~A. Rol, L.~Lao,
  K.~Bertels, and C.~G. Almudever, ``Openql : A portable quantum programming
  framework for quantum accelerators,'' 2020.

\bibitem{chapter4:unitarymatrices}
G.~D. Allen, ``Unitary matrices,'' in \emph{Lectures on Linear Algebra and
  Matrices}.\hskip 1em plus 0.5em minus 0.4em\relax College Station, TX: Texas
  A\&M University, 2003, ch.~4, pp. 157\--180.

\bibitem{art:genomesequencingzaid}
\BIBentryALTinterwordspacing
E.~J. Houtgast, V.-M. Sima, K.~Bertels, and Z.~Al-Ars, ``Hardware acceleration
  of bwa-mem genomic short read mapping for longer read lengths,''
  \emph{Computational Biology and Chemistry}, vol.~75, pp. 54 -- 64, 2018.
  [Online]. Available:
  \url{http://www.sciencedirect.com/science/article/pii/S1476927118301555}
\BIBentrySTDinterwordspacing

\bibitem{art:qibam}
A.~Sarkar, Z.~Al-Ars, C.~G. Almudever, and K.~Bertels, ``An algorithm for dna
  read alignment on quantum accelerators,'' 2019.

\bibitem{art:quantumassociativememory}
\BIBentryALTinterwordspacing
D.~Ventura and T.~Martinez, ``Quantum associative memory,'' \emph{Information
  Sciences}, vol. 124, no.~1, pp. 273 -- 296, 2000. [Online]. Available:
  \url{http://www.sciencedirect.com/science/article/pii/S0020025599001012}
\BIBentrySTDinterwordspacing

\bibitem{introductiontoliegroups}
A.~Savage, ``Introduction to {L}ie {G}roups,'' in \emph{Course notes of MAT
  1411/MAT 5158}.\hskip 1em plus 0.5em minus 0.4em\relax University of Ottawa,
  2015.

\bibitem{art:arbitrarytwoqubit}
\BIBentryALTinterwordspacing
S.~S. Bullock and I.~L. Markov, ``Arbitrary two-qubit computation in 23
  elementary gates,'' \emph{Physical Review A}, vol.~68, no.~1, Jul 2003.
  [Online]. Available: \url{http://dx.doi.org/10.1103/PhysRevA.68.012318}
\BIBentrySTDinterwordspacing

\bibitem{art:quantumcircuitsforgeneralmultiqubit}
M.~M\"ott\"onen, J.~J. Vartiainen, V.~Bergholm, and M.~M. Salomaa, ``Quantum
  circuits for general multiqubit gates,'' \emph{Phys. Rev. Lett.}, vol.~93, p.
  130502, Sep 2004.

\bibitem{art:entanglement20qubit}
\BIBentryALTinterwordspacing
G.~J. Mooney, C.~D. Hill, and L.~C.~L. Hollenberg, ``Entanglement in a 20-qubit
  superconducting quantum computer,'' \emph{Scientific Reports}, vol.~9, no.~1,
  Sep 2019. [Online]. Available:
  \url{http://dx.doi.org/10.1038/s41598-019-49805-7}
\BIBentrySTDinterwordspacing

\bibitem{art:minimaluniversal}
V.~V. Shende, I.~L. Markov, and S.~S. Bullock, ``Minimal universal two-qubit
  cnot-based circuits.''

\bibitem{synthesisofquantumlogiccircuits}
V.~Shende, S.~S.~Bullock, and I.~Markov, ``Synthesis of quantum logic
  circuits,'' \emph{Computer-Aided Design of Integrated Circuits and Systems,
  IEEE Transactions on}, vol.~25, pp. 1000 -- 1010, July 2006.

\bibitem{art:efficientdecompositionofquantumgates}
J.~Vartiainen, M{\"o}tt{\"o}nen, and M.~Salomaa, ``Efficient decomposition of
  quantum gates,'' \emph{Physical Review Letters}, vol.~92, no.~17, p. 177902,
  april 2004.

\bibitem{art:elementary}
A.~Barenco, C.~H. Bennett, R.~Cleve, D.~P. DiVincenzo, N.~Margolus, P.~Shor,
  T.~Sleator, J.~Smolin, and H.~Weinfurter, ``Elementary gates for quantum
  computation,'' 1995.

\bibitem{misc:rudimentaryquantumcompiler}
R.~R. Tucci, ``A rudimentary quantum compiler(2cnd ed.),'' 1999.

\bibitem{art:historyandgeneralityofthecsdecomposition}
C.~Paige and M.~Wei, ``History and generality of the cs decomposition,''
  \emph{Linear Algebra and its Applications}, vol. 208-209, pp. 303--326, 09
  1994.

\bibitem{misc:qubitergithub}
\BIBentryALTinterwordspacing
H.~Dekant, H.~Tregillus, R.~Tucci, and T.~Yin, ``Qubiter at github,'' 2020,
  accessed on: 23-10-2020. [Online]. Available:
  \url{https://github.com/artiste-qb-net/qubiter}
\BIBentrySTDinterwordspacing

\bibitem{art:decompositionofgeneralquantumgates}
M.~Möttönen and J.~Vartiainen, ``Decompositions of general quantum gates,''
  \emph{Frontiers in Artificial Intelligence and Applications}, 05 2005.

\bibitem{misc:eigendoc}
\BIBentryALTinterwordspacing
B.~J. (founder), G.~G. (guru), and many more, ``The eigen documentation,''
  2019, accessed on: 20-07-2020. [Online]. Available:
  \url{http://eigen.tuxfamily.org/index.php?title=Main_Page}
\BIBentrySTDinterwordspacing

\bibitem{art:openql2020}
N.~Khammassi, I.~Ashraf, J.~v.~Someren, R.~Nane, A.~M. Krol, M.~A. Rol, L.~Lao,
  K.~Bertels, and C.~G. Almudever, ``Openql : A portable quantum programming
  framework for quantum accelerators,'' 2020.

\bibitem{misc:lapackusersguide}
\BIBentryALTinterwordspacing
S.~Blackford, R.~Moore, and N.~Drakos, ``{LAPACK} users' guide,'' accessed on:
  23-10-2020. [Online]. Available:
  \url{http://www.netlib.org/lapack/lug/node71.html}
\BIBentrySTDinterwordspacing

\bibitem{art:realizationofathreequbitgate}
F.~Vatan and C.~P.~Williams, ``Realization of a general three-qubit quantum
  gate,'' 02 2004.

\bibitem{book:gelfand1989lectures}
\BIBentryALTinterwordspacing
I.~Gelfand, \emph{Lectures on Linear Algebra}, ser. Dover Books on Mathematics
  Series.\hskip 1em plus 0.5em minus 0.4em\relax Dover Publications, 1989.
  [Online]. Available: \url{https://books.google.nl/books?id=1ebLTz\_MtUcC}
\BIBentrySTDinterwordspacing

\end{thebibliography}

\appendix
\section{Tables}
\begin{table*}[h]
\centering
\setlength\tabcolsep{0.2em}
\renewcommand{\arraystretch}{1.1}
\caption{Total execution time at each line of \cref{lst:timingprogram} for the decomposition of unitary matrices of different sizes, in seconds.}
\label{tabexecutiontimes}
\begin{tabularx}{\linewidth}{|l|l|X|X|X|X|X|X|X|X|X|X|}
\hline
\rowcolor[HTML]{EFEFEF} 
Line                                                    & No decomp. & 1-qubit  & 2-qubit  & 3-qubit  & 4-qubit  & 5-qubit  & 6-qubit  & 7-qubit  & 8-qubit  & 9-qubit  & 10-qubit \\ \hline
Preamble                                               & 2.43E-3    & 2.46E-3 & 2.45E-3 & 2.24E-3 & 2.18E-3 & 2.50E-3 & 2.32E-3 & 2.35E-3 & 2.32E-3 & 2.75E-3 & 1.34E-2  \\ \hline
matrix = np.load(..) & 2.43E-3    & 7.96E-3 & 8.04E-3 & 7.87E-3 & 8.49E-3 & 8.20E-3 & 8.45E-3 & 7.84E-3 & 8.79E-3 & 8.75E-3 & 2.70E-2  \\ \hline
u1 = ql.Unitary("name",matrix)                     & 2.43E-3    & 7.99E-3 & 8.07E-3 & 7.91E-3 & 8.57E-3 & 8.46E-3 & 9.40E-3 & 1.13E-2 & 2.13E-2 & 4.86E-2 & 1.66E-1  \\ \hline
u1.decompose()                                         & 2.43E-3    & 8.15E-3 & 8.33E-3 & 8.30E-3 & 9.71E-3 & 1.61E-2 & 3.49E-2 & 0.12890 & 0.78170 & 4.59846 & 39.8166 \\ \hline
k.gate(u1, range(0,nqubits))                           & 2.56E-3    & 8.20E-3 & 8.39E-3 & 8.39E-3 & 9.95E-3 & 1.70E-2 & 3.63E-2 & 0.13497 & 0.80925 & 4.70238 & 40.2412 \\ \hline
p.add\_kernel(k)                                       & 2.56E-3    & 8.21E-3 & 8.39E-3 & 8.40E-3 & 9.97E-3 & 1.71E-2 & 3.63E-2 & 0.13550 & 0.81113 & 4.70981 & 40.2908 \\ \hline
compiler.compile(p)                                    & 8.49E-3    & 8.41E-3 & 8.60E-3 & 8.69E-3 & 1.05E-2 & 1.87E-2 & 4.00E-2 & 0.14867 & 0.87117 & 4.95417 & 41.3165 \\ \hline
\end{tabularx}
\end{table*}

\begin{table*}[h]
\centering
\setlength\tabcolsep{0.2em}
\renewcommand{\arraystretch}{1.1}
\caption{Additional memory allocated at each line of \cref{lst:timingprogram} for the decomposition of unitary matrices of different sizes, in MiB.}
\label{tab:memoryallocation}
\begin{tabularx}{\linewidth}{|l|l|X|X|X|X|X|X|X|X|X|X|}
\hline
\rowcolor[HTML]{EFEFEF} 
Line                           & 1-qubit & 2-qubit & 3-qubit & 4-qubit & 5-qubit & 6-qubit & 7-qubit & 8-qubit & 9-qubit & 10-qubit \\ \hline
Initial                       & 43.078  & 43.117  & 42.973  & 43.172  & 43.102  & 42.914  & 42.906  & 43.180  & 43.063  & 43.082   \\ \hline
matrix = np.load(…)            & 0       & 0       & 0       & 0       & 0       & 0       & 0       & 0.734   & 1.375   & 4.570    \\ \hline
u1 = ql.Unitary("name",matrix) & 0       & 0       & 0       & 0       & 0       & 0.766   & 1.855   & 3.258   & 12.160  & 48.141   \\ \hline
u1.decompose()                 & 0       & 0       & 0       & 0       & 0.820   & 0.867   & 1.945   & 5.750   & 12.156  & 46.184   \\ \hline
k.gate(u1, range(0,nqubits))   & 0       & 0       & 0       & 1.230   & 0.660   & 1.711   & 6.441   & 27.582  & 120.652 & 483.652  \\ \hline
p.add\_kernel(k)               & 0       & 0       & 0       & 0       & 0       & 0       & 0.316   & 1.344   & 4.441   & 18.105   \\ \hline
compiler.compile(p)            & 0       & 0       & 0       & 0       & 0.313   & 0.328   & 1.535   & 6.039   & 24.141  & 16.137   \\ \hline
\end{tabularx}
\end{table*}

\begin{table*}[h]
\centering
\setlength\tabcolsep{0.2em}
\renewcommand{\arraystretch}{1.1}
\caption{Aggregated execution times for the decomposition of unitary matrices of different sizes for OpenQL and Qubiter, in seconds.}
\label{tab:openqlqubitercomparison}
\begin{tabularx}{\linewidth}{|l|l|X|X|X|X|X|X|X|X|X|X|}
\hline
\rowcolor[HTML]{EFEFEF} 
                 & 1-qubit & 2-qubit & 3-qubit & 4-qubit           & 5-qubit           & 6-qubit              & 7-qubit              & 8-qubit              & 9-qubit              & 10-qubit             \\ \hline
OpenQL preamble  & 5.61E-4 & 8.36E-4 & 8.89E-4 & 8.03E-4 & 6.50E-4 & 5.38E-4 & 5.65E-4  & 6.29E-4 & 5.83E-4 & 9.56E-4  \\ \hline
OpenQL load      & 1.29E-3 & 1.81E-3 & 1.72E-3 & 1.66E-3 & 1.58E-3 & 1.39E-3 & 1.55E-3  & 1.81E-3 & 2.05E-3 & 9.01E-3  \\ \hline
OpenQL total     & 1.33E-3 & 2.06E-3 & 2.02E-3 & 2.81E-3 & 6.52E-3 & 2.39E-2 & 0.124    & 0.758   & 5.53    & 39.80    \\ \hline\hline
Qubiter preamble &         & 5.94E-6 & 7.84E-6 & 5.25E-6 & 5.39E-6 & 5.15E-6 & 1.03E-5  & 1.03E-5 & 1.05E-5 & 1.10E-5  \\ \hline
Qubiter load     &         & 8.27E-4 & 7.54E-4 & 8.71E-4 & 1.19E-3 & 1.68E-3 & 2.64E-03 & 5.11E-3 & 5.69E-3 & 7.59E-3  \\ \hline
Qubiter total    &         & 4.46E-3 & 9.83E-3 & 3.01E-2 & 0.162   & 0.496   & 2.15     & 9.22    & 37.07   & 178.41      \\ \hline
\end{tabularx}
\end{table*}

\begin{table*}[h]
\centering
\setlength\tabcolsep{0.2em}
\renewcommand{\arraystretch}{1.1}
\caption{Number of gates for OpenQL and Qubiter resulting from the decomposition of unitary matrices of different sizes, from counting lines in the generated assembly language files.}
\label{tab:openqlqubitergates}
\begin{tabularx}{\linewidth}{|l|X|X|X|X|X|X|X|X|X|X|l|}
\hline
\rowcolor[HTML]{EFEFEF} 
             & 1-qubit & 2-qubit & 3-qubit & 4-qubit & 5-qubit & 6-qubit & 7-qubit & 8-qubit & 9-qubit & 10-qubit &  Formula                                                  \\ \hline
OpenQL total           & 3       & 24      & 120     & 528     & 2208    & 9024    & 36480   & 146688  & 588288  & 2356224  & $\sfrac{9}{4}\cdot 4^n -3\cdot 2^n$    \\ \hline
OpenQL rotation gates  & 3       & 18      & 84      & 360     & 1488    & 6048    & 24384   & 97920   & 392448  & 1571328  & $\sfrac{3}{2}\cdot 4^n -\sfrac{3}{2}\cdot 2^n$     \\ \hline
OpenQL CNOTs           & 0       & 6       & 36      & 168     & 720     & 2976    & 12096   & 48768   & 195840  & 784896   & $\sfrac{3}{4}\cdot 4^n -\sfrac{3}{2}\cdot 2^n$     \\ \hline\hline
Qubiter total          & -        & 36      & 184     & 880     & 4064    & 18368   & 81792   & 360192  & 1572352 & 6814720  &  $\sfrac{3}{2}\cdot 4^n-\sfrac{1}{2}\cdot2^n$    \\ \hline
Qubiter rotation gates & -        & 22      & 92      & 376     & 1520    & 6112    & 24512   & 98176   & 392960  & 1572366  & $(\sfrac{3}{2} + \sfrac{1}{2}\cdot n)\cdot 4^n-2^n$       \\ \hline
Qubiter CNOTs          & -       & 14      & 92      & 504     & 2544    & 12256   & 57280   & 262016  & 1179392 & 5242354  & $\sfrac{1}{2}\cdot n\cdot4^n-\sfrac{1}{2}\cdot2^n$ \\ \hline
\end{tabularx}
\end{table*}

\begin{lstlisting}[label=lst:timingprogram,caption={Using unitary decomposition in OpenQL}]
import os
from openql import openql as ql
import numpy as np
import sys

nqubits = int(sys.argv[1])
		
ql.set_option('output_dir', os.path.join(curdir, 'test_new_output'))
ql.set_option('log_level', 'LOG_ERROR');
		
platf = ql.Platform("starmon", os.path.join(curdir, 'hardware_config_qx.json'))
program = ql.Program('example', platf, nqubits)
kernel = ql.Kernel("newKernel")
		
compiler = ql.Compiler('compiler1')
		
compiler.add_pass_alias("Writer","scheduledqasmwriter")
compiler.set_pass_option("scheduledqasmwriter", "write_qasm_files", "yes")
matrix = np.load('data/out_' + str(nqubits) + ".npy")
u1 = ql.Unitary("testname",matrix)
u1.decompose()	
kernel.gate(u1, range(0,nqubits))
program.add_kernel(kernel)
	
compiler.compile(program)
\end{lstlisting}

\end{document}